\RequirePackage{ifpdf}

\RequirePackage{fix-cm}

\documentclass[%
floatfix,
showkeys,
nofootinbib, %
superscriptaddress, %
]{revtex4-1}

\ifpdf
  \usepackage{cmap}
\fi
\usepackage{textcomp}

\usepackage[T1,T2A]{fontenc}

\usepackage[utf8]{inputenx}
\input{ix-utf8enc.dfu}

\usepackage[german,russian,english]{babel}

\usepackage{amsmath}
\usepackage{amssymb}

\usepackage{mathtools}
\mathtoolsset{
showonlyrefs,
mathic = true
}

\allowdisplaybreaks

\usepackage{hyperref}
\hypersetup{backref,
 colorlinks=false}
\hypersetup{pdfborder=0 0 0}

\usepackage{microtype}
\UseMicrotypeSet[protrusion]{alltext}

\usepackage{graphicx}

\usepackage[scanall]{psfrag}

\usepackage{listings}
\usepackage{listingsutf8}
\usepackage[frozencache]{minted}

\usepackage{textgreek}

\usepackage{fixltx2e}

\usepackage{nicefrac}

\makeatletter
\def\ps@pprintTitle{%
     \let\@oddhead\@empty
     \let\@evenhead\@empty
     \let\@oddfoot\@empty
     \let\@evenfoot\@oddfoot}
\makeatother

\usepackage{physics}
\usepackage{todonotes}

\usepackage{upgreek}
\usepackage{tensor}

\newcommand{\setR}{\mathbb{R}}

\DeclareMathOperator{\ldist}{d}

\DeclareMathOperator{\adist}{\delta}

\makeatletter
\@ifpackageloaded{upgreek}{%
\providecommand{\iiz}{\upvarepsilon}
}{%
\providecommand{\iiz}{\mathrm{\varepsilon}}
}
\makeatother

\begin{document}

\graphicspath{{image/dual-numbers/en/}{image/dual-numbers/}{image/}}

\title{Approaches to the implementation of generalized complex numbers in the Julia language}

\author{Migran N. Gevorkyan}
\email{gevorkyan-mn@rudn.ru}
\affiliation{Department of Applied Probability and Informatics,\\
  Peoples' Friendship University of Russia (RUDN University),\\
  6 Miklukho-Maklaya St, Moscow, 117198, Russian Federation}

\author{Anna V. Korolkova}
\email{korolkova-av@rudn.ru}
\affiliation{Department of Applied Probability and Informatics,\\
  Peoples' Friendship University of Russia (RUDN University),\\
  6 Miklukho-Maklaya St, Moscow, 117198, Russian Federation}

\author{Dmitry S. Kulyabov}
\email{kulyabov-ds@rudn.ru}
\affiliation{Department of Applied Probability and Informatics,\\
  Peoples' Friendship University of Russia (RUDN University),\\
  6 Miklukho-Maklaya St, Moscow, 117198, Russian Federation}
\affiliation{Laboratory of Information Technologies\\
  Joint Institute for Nuclear Research\\
  6 Joliot-Curie, Dubna, Moscow region, 141980, Russian Federation}

\begin{abstract}
  In problems of mathematical physics, to study the structures of
  spaces using the Cayley--Klein models in theoretical calculations,
  the use of generalized complex numbers is required. In the case of
  computational experiments, such tasks require their high-quality
  implementation in a programming language. The proposed small
  implementation of generalized complex numbers in modern programming
  languages have several disadvantages. In this article, we propose
  using the Julia language as the language for implementing
  generalized complex numbers, not least because it supports the
  multiple dispatch mechanism.

  The paper demonstrates the approach to the implementation of one of
  the types of generalized complex numbers, namely dual numbers. We
  place particular emphasis on the description of the use of the
  multiple dispatch mechanism to implement a new numerical type. The
  resulting implementation of dual numbers can be considered as a
  prototype for a complete software module for supporting generalized
  complex numbers.
\end{abstract}


\maketitle

\section{Introduction}
\label{sec:intro}

There is an approach that allows one to generalize the complex
numbers and get three different classes of generalized complex
numbers~\cite{yaglom:1964:projective-metrics::en,yaglom:complex-numbers-geometry::en,yaglom:1951:geometry-algebra::en,kulyabov:2020:jopcs:hyperbolic-numbers}.
In this approach, we set square equation
\begin{equation}
  \label{eq:x^2+px+q}
  z^2 + p z +q =0.
\end{equation}
with a determinant
\begin{equation}
  \label{eq:determinant}
  \Delta = p^2 -4 p q.
\end{equation}
Depending on the sign of the determinant, one gets the following results
are obtained complex numbers:
\begin{itemize}
\item $\Delta < 0$---elliptic (normal complex numbers);
\item $\Delta = 0$---parabolic (dual complex numbers);
\item $\Delta > 0$---hyperbolic (split complex numbers or double numbers).
\end{itemize}

Initially, these systems of complex numbers were introduced to describe
Cayley--Klein models~\cite{cayley:1859:sixth-memoir-upon-quantics,klein:1985:nicht-euklidische}.
However, they have other uses. In particular, dual
complex numbers can be used for problems of automatic
differentiation~\cite{L_Julia:2020:ForwardDiff}.

Existing implementations of generalized complex numbers are designed
for specific applications. For example, dual numbers are used for
automatic differentiation. The goal of this work is to create a pure
implementation of dual numbers that is as close as possible to their
mathematical definition. We use Julia programming language for this
task.

\subsection{Article structure}
\label{sec:structure}
The~\ref{sec:dual-complex} section provides sufficient description of
dual complex numbers. We use constructive definition, affecting only
operations with dual complex numbers. We try to avoid unnecessary
mathematical abstractions. In the section~\ref{sec:julia-features} we
give a General idea of the multiple dispatch mechanism --- the
fundamental idea of Julia language. Also, the multiple dispatch is the
basis for implementation of any user-defined type in Julia. In the
section~\ref{sec:dual-complex-julia} we describe in detail our
implementation of dual complex numbers in Julia.

\subsection{Notations and conventions}
\label{sec:notation}

  To denote a dual complex unit, we will use the symbol $\iiz$.

\section{Dual numbers}
\label{sec:dual-complex}

The dual number $z$ is defined algebraically as follows:
\begin{equation}
z = a + \iiz b,\; a,b\in\setR,\; \iiz^2 = 0, \iiz \neq 0.
\end{equation}
The value of $a$ will be called the real part, and $b$ ---
imaginary. There is no confusion in the terminology, because in this article we not use the ordinary complex numbers.

\subsection{The algebraic form}

Algebraic properties for two numbers $z_1 = a_1 + \iiz b_1$ и
$z_2 = a_2 + \iiz b_2$.
\begin{description}
  \item[Addition] $z_1 + z_2 = (a_1 + a_2) + \iiz (b_1 + b_2)$.
  \item[Subtraction] $z_1 - z_2 = (a_1 - a_2) + \iiz (b_1 - b_2)$.
  \item[Multiplication] $z_1 \cdot z_2 = (a_1 + \iiz b_1) \cdot (a_2 + \iiz b_2) = a_1 a_2 + \iiz (a_1 b_2 + b_1 a_2)$.
  \item[Conjunction] $\bar{z} = a - \iiz b$,
    $z \bar{z} = (a + \iiz b) \cdot (a - \iiz b) = a^2$.
  \item[Absolute value] $|z| = a$.
    The modulus of a dual number can be negative, and it can also be calculated by
the following formula:
    \begin{equation}
      |z| = \dfrac12 (z + \bar{z}) = \dfrac12 (a + \iiz
      b + a - \iiz b) = a.
    \end{equation}
  \item[Devision] The division of two dual numbers $z_1$ and $z_2$ is defined
for all $z_2$ such that $|z_2| \neq 0$:
  \begin{equation}
    \dfrac{z_1}{z_2} = \dfrac{a_1 + \iiz b_1}{a_2 + \iiz b_2} =
    \dfrac{(a_1 + \iiz b_1)(a_2 - \iiz b_2)}{(a_2 + \iiz b_2)(a_2 -
      \iiz b_2)} = \dfrac{a_1}{a_2} + \iiz \dfrac{b_1 a_2 - b_2
      a_1}{a^2_2}.    
  \end{equation}    
\end{description}

From the above properties, it can be seen that in imaginary parts of the numbers don't contribute to the real part of the result.

Consider a dual number of the form $c\iiz$, where $c \in \setR$. This is a dual number with zero absolute value. Such numbers have the following property:
\begin{equation}
  (c\iiz) \cdot (d\iiz) = cd\iiz^2 = 0,  
\end{equation}
from which it follows that for any number $c\iiz$ there is a number $d\iiz$ such that the product of these numbers is $0$. Such numbers are called \emph{zero divisors}.

\subsection{Trigonometric form}

A dual number $z$ such that $|z| \neq 0$ can be written as follows:
\begin{equation}
a + \iiz b = a \Big( 1 + \dfrac{b}{a}\iiz \Big) = r(1 + \varphi \iiz),
\end{equation}
where the values $r = |z| = a$ and $\iiz = Arg z = b/a$ are called \emph{module} and \emph{argument} of the dual number $z$, respectively.

This form of a dual number is some analog of the trigonometric form of an ordinary complex number and we will continue to call it \emph{trigonometric form} of the dual number.

For the conjugate number $\bar{z}$, the trigonometric form is:
\begin{equation}
\bar{z} = r(1 - \varphi \iiz),
\end{equation}
In this case, $|\bar{z}| = |z|$ is an absolute value, and $Arg\bar{z} = - b/a$ is an  argument of a dual number in trigonometric form.

The analogy with the trigonometric form of a complex number continues further when considering multiplication and division. When multiplying two dual numbers $z_1$ and $z_2$, we get
\begin{equation}
z = r_1(1 + \varphi_1 \iiz) \cdot r_2(1 + \varphi_2 \iiz) = r_1 r_2 \Big(1 + (\varphi_1 + \varphi_2)\iiz\Big),
\end{equation}
in other words, when multiplying, the arguments are added and the modules are multiplied.

In the case of division, we get
\begin{equation}
\dfrac{z_1}{z_2} = \dfrac{r_1(1+\varphi_1 \iiz_1)}{r_2(1 + \varphi_2\iiz)} = \dfrac{r_1 (1 + \varphi_1 \iiz) r_2 (1 - \varphi_2\iiz)}{r_2(1 + \varphi_2\iiz)r_2(1 - \varphi_2\iiz)} = \dfrac{r_1(1 + (\varphi_1 - \varphi_2)\iiz)}{r_2 (1 - \varphi_2 \iiz + \varphi_2 \iiz)} = \dfrac{r_1}{r_2}(1 + (\varphi_1 - \varphi_2)\iiz),
\end{equation}
in other words, when dividing, arguments are subtracted, and modules are divided.

Note that in the case of division, the number $z_2$ must have a non-zero module $|z_2| \neq 0$.

In trigonometric form, the operation of raising to the natural power of $n$ looks especially simple:
\begin{equation}
z^n = \big(r (1 + \varphi \iiz)\big)^n = \underbrace{r\cdot r \cdot \ldots \cdot r}_{n}\big(1 + \underbrace{(\varphi + \varphi + \ldots + \varphi)}_{n}\big) = r^n (1 + n\varphi\iiz) = a^n + \iiz n a^{n-1}b.
\end{equation}

For finding
$\sqrt[n]{z} = \sqrt[n]{r (1 + \varphi\iiz)}$, $n \in \mathbb{N}$, assume that
$\sqrt[n]{z} = z_1$, then by definition $z^n_1 = z$, hence
\begin{equation}
r_1^n(1 + n\varphi_1\iiz) = z = r(1 + \varphi\iiz),
\end{equation}
what gives
\begin{equation}
r_1 = \sqrt[n]{r}, \varphi_1 = \dfrac{\varphi}{n}, \sqrt[n]{z} = \sqrt[n]{r}\left(1 + \dfrac{\varphi}{n}\iiz\right).
\end{equation}

Note that for an odd $n$, the root $\sqrt[n]{z}$ always exists, and for an even $n$ --- only for the number $z$ with a non-negative module $|z| = r \neq 0$. In algebraic form, the formula for the root is:
\begin{equation}
\sqrt[n]{a + b\iiz} = \sqrt[n]{a} + \dfrac{b a^{\frac{1-n}{n}}}{n}\iiz.
\end{equation}
\subsection{Matrix form}

All algebraic operations on dual numbers can be reduced to matrix operations if we consider:
\begin{equation}
\iiz \leftrightarrow
\begin{pmatrix}
0 & 1 \\
0 & 0
\end{pmatrix}
\;\;
z = a + b\iiz \leftrightarrow
\begin{pmatrix}
a & b \\
0 & a
\end{pmatrix}.
\end{equation}

Then, for example:
\begin{equation}
z_1 \cdot z_2 \leftrightarrow
\begin{pmatrix}
a_1 & b_1 \\
0 & a_1
\end{pmatrix}
\begin{pmatrix}
a_2 & b_2 \\
0 & a_2
\end{pmatrix}
=
\begin{pmatrix}
a_1a_2 & a_1b_2 + a_2 b_1 \\
0 & a_1a_2
\end{pmatrix}
\leftrightarrow
a_1a_2 + (a_1b_2 + a_2 b_1)\iiz.
\end{equation}

If the programming language supports vectorization, then in theory the matrix form of dual numbers can give you a performance gain. However, the effect is unlikely to be significant in practice.

\subsection{Taylor series expansion}

We can use
$\iiz^2 = \iiz^3 = \ldots = \iiz^n = 0$ $\forall n \in \mathbb{N}$, to get:
\begin{equation}
\exp(b\iiz) = \sum\limits^{\infty}_{n=0}\dfrac{(b\iiz)^n}{n!} = 1 + b\iiz + \dfrac{b^2\iiz^2}{2!} + \ldots = 1 + b\iiz,
\end{equation}
\begin{equation}
\exp(a + b\iiz) = e^ae^{b\iiz} = e^a(1 + b\iiz).
\end{equation}

A more general formula is derived from the Taylor series for the function $f(z)$ at the point $a$:
\begin{equation}
f(a + \iiz b) = \sum\limits^{\infty}_{n=0}\dfrac{f^{(n)}(a)(a + \iiz b - a)^n}{n!} = \sum\limits^{\infty}_{n=0}\dfrac{f^{(n)}(a)\iiz^nb^n}{n!} = f(a) + f'(a)b\iiz + \dfrac{f^{''}(a)\iiz^2b^2}{2!} + \ldots = f(a) + f'(a)b\iiz.
\end{equation}
As a result, we get an extremely important equation:
\begin{equation}
f(a + \iiz b) = f(a) + f'(a)b\iiz,
\end{equation}
which gives a way to calculate the values of functions from a dual number, if the value of the derivative $f'(a)$ is known. On the other hand, the same formula allows one to calculate the value of the first derivative at the point $a$.

\subsection{Elementary functions of dual numbers}

The formula $f(a + \iiz b) = f(a) + f'(a) b\iiz$ allows you to extend elementary functions to the set of dual numbers, since the right part of the formula contains only the values of the function $f$ from the real number $a$.

Here is a brief summary of basic elementary functions for illustration.

\begin{center}
  {\renewcommand{\arraystretch}{1.2}%
  \begin{tabular}{l|l}
    Trigonometric function & Inverse trigonometric functions \\ \hline
    $\sin(a+\iiz b) = \sin{a} + b \iiz\cos{a}$ & $\arcsin(a + \iiz b) = \arcsin{a} + b\iiz / \sqrt{1 - a^2}$\\
    $\cos(a+\iiz b) = \cos{a} - b \iiz\sin{a}$ & $\arccos(a + \iiz b) = \arccos{a} - b\iiz / \sqrt{1 - a^2}$\\
    $\tg(a+\iiz b) = \tg{a} + b \iiz / \cos^2{a}$ & $\arctg(a + \iiz b) = \arctg{a} + b\iiz / (1 + a^2)$\\
    $\ctg(a+\iiz b) = \ctg{a} - b \iiz / \sin^2{a}$ & $\arcctg(a + \iiz b) = \arctg{a} - b\iiz / (1 + a^2)$
  \end{tabular}}
\end{center}

\begin{center}
  {\renewcommand{\arraystretch}{1.2}%
  \begin{tabular}{l|l}
    Power functions & Logarithmic functions and exponent\\ \hline
    $(a + \iiz b)^n = a^n + n a^{n-1} b \iiz$ & $\exp(a + \iiz b) = \exp{a} + b\iiz \exp{a}$\\
    $\sqrt[n]{a + \iiz} = \sqrt[n]{a}\Big( 1 + \dfrac{b\iiz}{n a} \Big)$ & $\log_{c}(a + \iiz b) = \log_c{a} + b\iiz / a \ln{a}$
  \end{tabular}}
\end{center}

\subsection{Calculating the first derivative of a real value function using dual numbers}

The formula $f(a + \iiz b) = f(a) + f'(a) b\iiz$ can also be used to calculate the value of the first derivative of the real function $f(x)$ at the point $a$, if the value of the function $f(a + \iiz b)$is known. From the point of view of analytical calculations, this method of finding the derivative does not make sense, since the value $f (a + \iiz b)$ is obtained analytically from the same formula. However, it provides a numerical method for obtaining the value of the first derivative using numerical calculations without any additional error. This method of finding the derivative with help of dual numbers is called \emph{automatic differentiation}. It also should mensioned that automatic differentiation using dual numbers are limited to derivatives of the first order.

Let's first look at a simple example of automatic differentiation, and then discuss the general implementation principle for programming languages.

As an example, let's find the derivative of the function $f(x) = x\sin{x}$:
\begin{equation}
f(a + \iiz b) = (a + \iiz b) \sin(a + \iiz b).
\end{equation}

Knowing the value of $\sin(a + \iiz b) = \sin{a} + b\iiz\cos{a}$, it is easy to calculate $f (a + \iiz b)$:
\begin{equation}
f(a + \iiz b) = (a + \iiz)(\sin{a} + b\iiz\cos{a}) = a\sin{a} + (\sin{a} + a\cos{a})b\iiz.
\end{equation}

Comparing with the general formula $f(a + \iiz b) = f(a) + f'(a) b\iiz$, we get the value $f'(a) = \sin{a} + b\iiz\cos{a}$.

In general, to find $f'(a)$, it is enough to find the value of $f'(a + \iiz b)$, take the imaginary part of this dual number and divide it by $b$. In addition, since the choice of the number $b$ is arbitrary, we can choose it equal to $1$ and get rid of the need to divide by $b$.

To apply automatic differentiation using dual numbers, we need the programming language which allows user-defined types, as well as overloading arithmetic operations and elementary functions for this type. This is especially easy for object-oriented languages and languages that support function overloading or multiple dispatching.

\section{Dynamical dispatch in Julia language}

\label{sec:julia-features}

  The language Julia~\cite{L_Julia:2020:site} appeared relatively
  recently, but has already gained popularity as a language for
  scientific computing. We will assume that readers are already
  familiar with this language, and briefly focus only on the concept
  of \emph{multiple
    dispatching}~\cite{bezanson:2018:julia-dinamism-by-design,zappa:2018:julia-subtyping,bezanson:2017:julia-numeric-computing,bezanson:2012:julia-dinamic-language},
  which is the basis of the language, its understanding is essential
  for further presentation.

  \emph{Dynamic dispatch} is a mechanism that allows to select the
  specific implementation of a polymorphic function or operator from a
  set and call it in a specific case.

  \emph{Multiple dispatching} is based on dynamic dispatching. In this
  case, the choice of exact implementation of polymorphic function is
  made based on the type, number, and order of the function's
  arguments. This is the runtime polymorphic dispatch. In addition to
  the term <<multiple dispatching>>, the term \emph{multimethod} is
  also used.

  The mechanism of multiple dispatch is similar to the mechanism of
  functions and operators overload, implemented, for example, in the
  C++ language. Function overloading, however, is performed
  exclusively at the compilation stage, whereas multiple dispatching
  must also work at the program run-time (run-time polymorphism).

  Julia supports overloading functions at the compilation time if all
  data types used in the function can be casted at the compilation
  stage (so called \emph{type stable} functions). The JIT compiler
  creates efficient implementations for each combination of argument
  types.

  If it is impossible to cast data types at the compilation stage
  (type unstable function), the dynamic dispatching mechanism is
  enabled. The compiler will not be able to create a specialized
  version, but will create a generic version that runs slowly.

  This approach allows one to combine the speed of a compiled language
  with strict static typing with the flexibility of an interpreted
  language with dynamic typing.

\section{Dual numbers implementation}
\label{sec:dual-complex-julia}

\subsection{Existing implementations of dual numbers in Julia}

The authors are familiar at least with two realizations of dual numbers in
Julia language:
\begin{itemize}
  \item type \verb|Dual| in the module for automatic differentiation Forward\-Diff~\cite{L_Julia:2020:ForwardDiff};
  \item separate module Dual\-Numbers~\cite{L_Julia:2020:DualNumbers}, the development of which is frozen in favor of ForwardDiff.
\end{itemize}

In this section we will describe our proposed implementation of dual numbers in the Julia language. It is based on a built-in type of complex numbers \verb|Complex|~\cite{L_Julia:repo:2020}, and on the Dual\-Numbers module. The proposed implementation in this paper serves only as a example of creating a custom data type for the Julia language.

We put clarity of presentation at the first place, so many computational optimizations were deliberately omitted in favor of a larger clarity. For example, in the Dual\-Numbers module, the dual number $z = a + \iiz b$ can have ordinary complex number components $a$ and $b$. Also in Dual\-Numbers for implementing elementary functions $f(z)$ from dual numbers $z$ a third-party module is used, which defines differentiation rules for functions from the real variable. This allows to automate the definition of the $f(z)$ function by the formula $f(a + \iiz b) = f(a) + f'(a)b\iiz$, since it is not necessary to explicitly write out derivatives of $f'(a)$.

In our implementation, $a$ and $b$ are only real values and elementary functions are defined explicitly.

\subsection{Data structure}

\begin{figure}[!h]
  \centering
  \includegraphics[width=\linewidth]{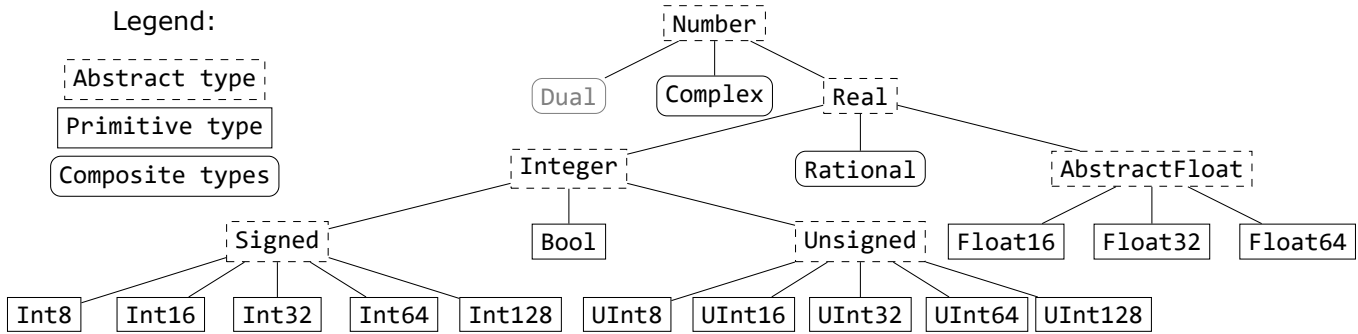}
  \caption{Our Dual type in Julia built-in types hierarchy}
  \label{fig:01}
\end{figure}

Adding any custom data type in Julia starts with defining the data structure. For dual numbers we define the structure
\verb|Dual|:
\begin{minted}{julia}
  struct Dual{T<:Real} <: Number
    "real part"
    x::T
    "imaginary part"
    y::T
  end
\end{minted}

The structure contains only two fields: the real \verb|x| and the imaginary \verb|y| parts of the dual number. Both fields in the structure must have the same parametric type \verb|T|, which must be a subtype of the abstract type \mintinline{julia}{Real}, as indicated by the \mintinline{julia}{<:} operator. The \verb|Dual| type itself  is a subtype of the abstract type \mintinline{julia}{Number}. In this way, the dual numbers are embedded in an existing hierarchy of types (see figure~\ref{fig:01}).

Immediately after declaring a new structure, we can create variables of the Dual type using the default constructors. We must specify both fields of the structure as arguments of the Dual function:
\begin{minted}{julia}
z = Dual(1, 2) # the same type T of the arguments
Dual{Float64}(1, 2.2) # casting arguments to the same type
\end{minted}

If the argument type is the same, we don't need to explicitly specify the \verb|T| parameter. If the arguments have different types, the parametric type must be specified. In our example, both arguments are casted to the \mintinline{julia}{Float64} type.

  \subsection{Overloading functions in the show example}

To print the values of variables of the Dual type we must to overload the \verb|show|function. This function is intended for formatted printing of data to standard output and to REPL. The \verb|write|, \verb|print|, and \verb|println| functions are used to output a minimal text representation of data. The last two functions print data only to the standard output stream. For simple data, one can make no difference between formatted and compact views and overload only \verb|show| function. Then the other functions will call \verb|show|.

In terms of multiple dispatch, overloading a function means adding a new \emph{method} to the function. To add a method to the \verb|show| function, we define it as follows:
\begin{minted}{julia}
Base.show(io::IO, z::Dual) = show(io, z.x, " + ", z.y, "ε")
\end{minted}
The short syntax for function definition is used here. The prefix \verb|Base.| is needed, since the standard methods of the \verb|show| function are located in the \verb|Base| module. Functions from this module can be called directly without a prefix, but for overloading the \verb|Base| prefix is essential.

It is worth noting that our version of the \verb|show| method is simplified, since it does not take into account special cases, for example, the negative imaginary part.

Next, we will add methods for many functions from \verb|Base|, so that our \verb|Dual| type becomes sufficiently functional.

  \subsection{Additional constructors}

Let us create some additional constructors. To do this, we should overload the default constructor \verb|Dual|.

In total we will set 5 additional constructors:

\begin{minted}{julia}
Dual(x::Real, y::Real)= Dual(promote(x, y)...) # 1
Dual{T}(x::Real) where {T<:Real} = Dual{T}(x, 0) # 2
Dual(x::Real) = Dual(promote(x, 0)...) # 3
Dual{T}(z::Dual) where {T<:Real} = Dual{T}(z.x, z.y) # 4
Dual(z::Dual) = Dual(promote(z.x, z.y)...) # 5
\end{minted}

  The first constructor allows us not to specify the \verb|T| parameter every time if the arguments have different types. The \mintinline{julia}{promote} function converts the type of arguments passed to it to a common type and returns the result as a tuple. The Postfix operator \verb|...| unpacks the tuple and passes its elements as arguments to the constructor function. The language core defines casting rules for all subtypes of the abstract type \mintinline{julia}|Real|, so now the constructor will work correctly for any combination of arguments, as long as the \mintinline{julia}|T<:Real| is true. For example, the following code will work correctly:
\begin{minted}{julia}
  Dual(1//2, π) # 0.5 + π*ε
\end{minted}

We passed a rational number (type \mintinline{julia}|Rational|) and a built-in global constant (number $\pi$) of the type \mintinline{julia}|Float64| to the constructor. After that, the type casting rule worked and both arguments were converted to the more general type \mintinline{julia}|Float64|.

The second and third additional constructors allow one to omit the imaginary part if it is equal to zero:
\begin{minted}{julia}
Dual{Float32}(1) # 1.0 + 0ε, 2 constructor
Dual(1//2) # 1//2 + 0ε, 3 constructor
\end{minted}

Constructor number 2 is a parametric function that is declared using the \mintinline{julia}|where| construct. The \verb|T| parameter is a subtype of the abstract type \mintinline{julia}|Real|. Constructor number 3 works similarly to first constructor. The fourth and fifth constructors allow one to pass in an argument to the constructor other dual number.

For more convenience, we can also create a separate constant for the imaginary unit $\iiz$:
\begin{minted}{julia}
const ε = Dual(0, 1)
\end{minted}

After overloading arithmetic operations, this constant will allow us to create new dual numbers using an expression as close as possible to their algebraic notation:
\begin{minted}{julia}
z = 1 + 2ε
\end{minted}

  \subsection{Access to structure fields}

Structures in Julia are immutable by default, that is, once we create the variable \verb|z|, we can access the fields of the structure itself using the point operator, but we cannot modify the value of these fields:
\begin{minted}{julia}
  @show z.x z. y # We can read the field value
  z.x = 1 # Error! We can't change it.
\end{minted}

To create mutable data types, the \mintinline{julia}{mutable struct} structure is provided. However, for our example, an immutable structure is more appropriate, since the requirement for mutability imposes performance restrictions, which is undesirable for a numeric type.

Julia doesn't have access modifiers for fields like \verb|public| or \verb|private|, so structure fields are always readable. However, it is considered a good programming style to encapsulate structure fields and provide an interface for accessing them. This style is used, for example, for the built-in type of complex numbers.

There are two ways to encapsulate fields. The first method is to create special interface functions for accessing fields. In our case it is sufficient to define the following functions:

\begin{minted}{julia}
  Base.real(z::Dual) = z.x
  Base.imag(z::Dual) = z.y
  Base.reim(z::Dual) = (z.x, z.y)
\end{minted}

The same functions are defined in the built-in Complex module. Next, we should call only these functions everywhere to get the structure fields, instead of accessing the fields by name directly. This will allow developers to refactor the structure in the future, for example, rename or add new fields. Backward compatibility is easy to maintain by rewriting only the interface functions. Performance will not suffer, since the JIT compiler will replace calls to these functions directly with their code, since they are extremely simple (inline functions).

The second approach is to use the \mintinline{julia}{getproperty} function, which has been available since version 0.7 of the Julia language. Overloading this function allows us to set additional names for accessing fields in the structure. So, if we write the following method for example:
\begin{minted}{julia}
function Base.getproperty(z::Dual, p::Symbol)
  if p in (:real, :a, :r) # real part aliases
    return getfield(z, :x)
  elseif p in (:imag, :b, :i)  # imaginary part aliases
    return getfield(z, :y)
  else
    return getfield(z, p)
  end
end
\end{minted}
then we will be able to access the real and imaginary parts of the number $z$ in four different ways:
\begin{minted}{julia}
z.x == z.a == z.r == z.real # returns true
z.y == z.b == z.i == z.imag # returns true
\end{minted}

This approach is more flexible, since when changing the structure for backward compatibility, it is enough to modify only one method \mintinline{julia}{getproperty}, not all the interface functions (if the structure is complex it can by lots of such functions). In addition, the programmer can use any name from the aliases list to accessing the structure fields.

For the Dual type, we applied both approaches, since the presence of the functions \verb|real| --- equivalent to $\Re(z)$, and \verb|imag| --- equivalent to $\Im(z)$ is mathematically justified.

  \subsection{Unary functions}

The \verb|Base| module defines the functions \verb|one| and \verb|zero|, which return a multiplication identity element and an addition identity element, respectively. In the case of dual numbers, the multiplication identity is the real unit, and the addition identity is the real zero.

For the Dual type, we define the following single-line methods:

\begin{minted}{julia}
  Base.one(::Type{Dual{T}}) where T <: Real = Dual(one(T))
  Base.one(z::Dual) = Dual(one(z.x))
  
  Base.zero(::Type{Dual{T}}) where T <: Real = Dual(zero(T))
  Base.zero(z::Dual) = Dual(zero(z.x))
\end{minted}

The first version of the \verb|one| and \verb|zero| functions takes the data type as an argument and returns one or zero of this type, respectively. Since for dual numbers it is $1$ and $0$ from $\mathbb{R}$, it is sufficient to return a dual number with a zero imaginary part. The real part will be $1$ and $0$ of the parametric type \verb|T|. For all standard types \verb|T<:Real| the \verb|one| and \verb|zero| methods are defined in the \verb|Base| module, which we used.

The second variant of functions takes a specific object of the \verb|Dual| type as an argument and also returns a identity using methods from the \verb|Base| module.

For complex numbers in \verb|Base|, the conjugation functions \verb|conj|, the absolute value \verb|abs|, and the argument \verb|arg| are defined:

\begin{minted}{julia}
  Base.conj(z::Dual) = Dual(z.x, -z.y) 
  Base.abs(z::Dual)  = abs(z.x)
  Base.abs2(z::Dual) = z.x^2
  function Base.arg(z::Dual)
    @assert !iszero(z.x)
    return z.y / z.x
  end
\end{minted}

For the \verb|arg| method, we have provided a check for the inequality of the real part of the dual number to zero, since otherwise we will get a division by zero. Using the standard function \verb|iszero| allows us not to worry about accounting for errors in the representation of real numbers using floating-point numbers. The \mintinline{julia}{@assert} macro throws an exception \mintinline{julia}{AssertionError} if the actual part is equal to zero.

The \verb|inv| function defines the inverse number for this number \verb|z|:
\begin{minted}{julia}
  function Base.inv(z::Dual)
    @assert !iszero(z.x)
    return Dual(1/z.x, -z.y/z.x^2)
  end
\end{minted}
There should also be an exception for the null real part.

  \subsection{Comparison function}

The \verb|Base| module also defines a number of functions that return the truth, if a particular condition is true. List some of these functons are:
\begin{itemize}
  \item \verb|isreal| is real number;
  \item \verb|isinteger| is integer;
  \item \verb|isfinite| the number is finite;
  \item \verb|isnan| the argument has the type \mintinline{julia}{NaN};
  \item \verb|isinf| the argument has the type \mintinline{julia}{Inf};
  \item \verb|iszero| the number is an addition identity (zero);
  \item \verb|isone| the number is a multiplication identity (one).
\end{itemize}

Methods for these functions for the \verb|Dual| one-to-one case
repeats methods for the built-in \verb|Complex| type, so here we
do not provide their source code.

In addition to unary functions for dual numbers, it makes sense to implement methods for the comparison operator \verb|==|. Operators in Julia are no different from functions, and adding methods for them is exactly the same. The only difference is that we need to add the character \verb|:| after \verb|Base.|, and since the \verb|==| operator consists of two characters, we must frame it with parentheses:
\begin{minted}{julia}
Base.:(==)(z::Dual, u::Dual) = (real(z) == real(u)) && (imag(z) == imag(u))
\end{minted}

It makes sense to compare dual numbers with real numbers as well, if the dual number has a zero imaginary part. In order for the operator to commute for arguments of different types, we must define two methods:

\begin{minted}{julia}
Base.:(==)(z::Dual, x::Real) = isreal(z) && real(z) == x
Base.:(==)(x::Real, z::Dual) = isreal(z) && real(z) == x
\end{minted}

  \subsection{Arithmetic operations}

We also need to overload arithmetic operators such as the unary operators \verb|+| and \verb|-| and the binary operators \verb|+|, \verb| -|, \verb|*| and \verb|/|. The implementation of the \verb|+|, \verb| -|, and \verb|*| operators is trivial:

\begin{minted}{julia}
  Base.:+(z::Dual) = Dual(+z.x, +z.y)
  Base.:-(z::Dual) = Dual(-z.x, -z.y)
  
  Base.:+(z::Dual, u::Dual) = Dual(z.x + u.x, z.y + u.y)
  Base.:-(z::Dual, u::Dual) = Dual(z.x - u.x, z.y - u.y)
  
  Base.:*(z::Dual, u::Dual) = Dual(z.x * u.x, z.x * u.y + z.y * u.x)
\end{minted}

In the case of the \verb|/| operator, we should take into account the equality of the divisor to zero:
\begin{minted}{julia}
  function Base.:/(z::Dual, u::Dual)
    @assert !iszero(u.x)
    return Dual(z.x/u.x, (z.y*u.x-u.y*z.x)/u.x^2)
  end
\end{minted}

For binary operators, we do not need to implement separate cases of arguments of different types, because in this case the promotion mechanism to a common type will call \verb|promote| function automatically.

  \subsection{Types promotion}

To make type promotion mechanism to work, we must define the \emph{type promotion} rules. Without these rules, for example, the next operation fails with an error:

\begin{minted}{julia}
  >>Dual(1, 2) + 2
  ERROR: promotion of types Dual{Int64} and Int64 failed to change any arguments
\end{minted}

This error occurs because there is no rule that can be used to determine the common type for numbers of the type \mintinline{julia}{Dual{Int64}} and the type \mintinline{julia}{Int64}.

To define such a rule, we have to add a method for the \verb|promote_rule| function from the \verb|Base| module. Any number of the \mintinline{julia}{Real} type can participate in the arithmetic operation with a dual number, since this number can be interpreted as a dual with a zero imaginary part:
\begin{minted}{julia}
Base.promote_rule(::Type{Dual{T}}, ::Type{S}) where {T<:Real, S<:Real} = Dual{promote_type(T, S)}
\end{minted}

We should also provide a rule that will work for operators with two dual numbers with different parametric types \verb|T| and \verb|S|. In this case, we need to find a common type for the \verb|T| and \verb|S| types:
\begin{minted}{julia}
Base.promote_rule(::Type{Dual{T}}, ::Type{Dual{S}}) where {T<:Real, S<:Real} = Dual{promote_type(T, S)}
\end{minted}

  \subsection{Raising to a rational degree}

To raise a number to a power, the \verb|^| operator is used, which should also be overloaded for integer and rational powers. In the case of an integer degree, we use the formula
\begin{equation}
(a + \iiz b)^n = a^n + nba^{n-1}\iiz.
\end{equation}

There are several special cases to consider:
\begin{itemize}
\item if $n=0$, then $z^n = 1$;
\item if $n=1$, then $z^n = z$;
\item if $n > 0$, the formula works for any $a$ and $b$;
\item if $n < 0$, the condition $a \neq 0$must be met.
\end{itemize}

If we consider all these cases, we get the following implementation:
\begin{minted}{julia}
  function Base.:^(z::Dual, n::Integer)::Dual
    x, y = reim(z)
    if n == 0
      return one(z)
    elseif n == 1
      return z
    elseif n > 1
      return Dual(x^n, n*y*x^(n-1))
    else #n < 0
      if iszero(x)
        throw(DomainError(:x, "negative exponentiation is only defined for real(z)!= 0"))
      else
        return Dual(x^n, n*y*x^(n-1))
      end
    end
  end
\end{minted}

If the real part of the dual number is zero (\verb|iszero(x)|) and $n<0$, the function throws an exception \mintinline{julia}{DomainError} (out of the range of acceptable values).

For a rational degree, a more complex formula is used:
\begin{equation}
(a + \iiz b)^{\frac{n}{m}} = a^{\frac{n}{m}}
\qty(1 + \iiz\frac{n}{m}\frac{b}{a})
= a^{\frac{n}{m}} +
\iiz\frac{n}{m}ba^{\frac{n}{m} - 1},
\end{equation}
for which the cases of even and odd $m$ should be provided. For an odd $m$, the range of acceptable values includes $a < 0$, and for an even $m$, you should limit yourself to $a \geqslant 0$:
\begin{minted}{julia}
  function Base.:^(z::Dual, q::Rational)::Dual
    x, y = reim(z)
    n, d = numerator(q), denominator(q)
    if n == 0
      return one(z)
    elseif isodd(d)
      return Dual(x^q, q*y*x^(q-1))
    else
      if x < 0
        throw(DomainError(:x, "even radical for dual number z is only defined for real(z)>=0"))
      else
        return Dual(x^q, q*y*x^(q-1))
      end
    end
  end
\end{minted}

It makes sense to separately overload the functions for square and cubic roots \verb|sqrt| and \verb|cbrt|:
\begin{minted}{julia}
  function Base.sqrt(z::Dual)::Dual
    x, y = reim(z)
    if x < 0
      throw(DomainError(:x, "sqrt for dual number z is only defined for real(z)>=0"))
    else
      return Dual(√x, y/(2*√x))
    end
  end
  
  Base.cbrt(z::Dual) = Dual(cbrt(z.x), z.y*cbrt(z.x)/(3z.x))
\end{minted}

  \subsection{Elementary functions}

Elementary functions are calculated using the formula $f(a + \iiz b) = f(a) + f'(a) b\iiz$. Note that many of them are not defined for the case of the zero real part of the number $z = a + \iiz b$:
\begin{minted}{julia}
  function Base.exp(z::Dual)
    e = exp(z.x)
    return Dual(e, e * z.y)
  end

  Base.log(z::Dual) = Dual(log(z.x), z.y/z.x)
  Base.log(b, z::Dual) = Dual(log(b, z.x), (z.y/z.x) * log(z.x))

  #======== Trigonometric ========#

  Base.sin(z::Dual) = Dual(sin(z.x), +z.y*cos(z.x))
  Base.cos(z::Dual) = Dual(cos(z.x), -z.y*sin(z.x))

  Base.tan(z::Dual) = Dual(tan(z.x),  z.y/cos(z.x)^2)
  Base.cot(z::Dual) = Dual(cot(z.x), -z.y/sin(z.x)^2)

  Base.asin(z::Dual) = Dual(asin(z.x),  z.y/sqrt(1-z.x^2))
  Base.acos(z::Dual) = Dual(acos(z.x), -z.y/sqrt(1-z.x^2))

  Base.atan(z::Dual) = Dual(atan(z.x),  z.y/(1+z.x^2))
  Base.acot(z::Dual) = Dual(acot(z.x), -z.y/(1+z.x^2))

  #======== Hyperbolic ========#

  Base.sinh(z::Dual) = Dual(sinh(z.x), z.y*cosh(z.x))
  Base.cosh(z::Dual) = Dual(cosh(z.x), z.y*sinh(z.x))

  Base.tanh(z::Dual) = Dual(tanh(z.x), z.y/cosh(z.x)^2)
  Base.coth(z::Dual) = Dual(coth(z.x), z.y/sinh(z.x)^2)
\end{minted}
\section{Results}
\label{sec:results}

  The paper describes the preliminary implementation of dual complex
  numbers and basic operations on them in the Julia language.

\section{Discussion}
\label{sec:discussion}

The implementation of dual complex numbers on Julia is completely based on the mechanism of multiple dispatching. Thus, we not only implemented a certain set of operations on dual numbers, but also demonstrated the power of this mechanism.

It should also be noted that in contrast to the implementation of the \verb|Dual| type in the automatic differentiation package \verb|ForwardDiff|~\cite{L_Julia:2020:ForwardDiff}, our proposed implementation is cleaner. For example, in the above package, an ordinary complex number can be used as a coefficient before a dual complex unit. It is clear that this is due to the specifics of using dual numbers in this package for automatic differentiation. But this type of number is more likely to belong to quaternions~\cite{hamilton:1866:elements-quaternions,hamilton:1866:elements-quaternions,yefremov:2005:quaternions-biquaternions}, rather than the proper complex numbers.

\section{Conclusion}
\label{sec:conclusion}

In this paper, a prototype of the implementation of generalized complex numbers in the Julia language was made, namely, the implementation of dual complex numbers. Having a multiple dispatching mechanism in the Julia language makes it very easy to implement new numeric types within the existing programming language infrastructure. We propose to further extend this prototype to more general implementation of generalized complex numbers and generalized quaternions.

\begin{acknowledgments}

  The publication has been prepared with the support of the
  Russian Foundation for Basic Research (RFBR) according to the research project
  No~19-01-00645.

\end{acknowledgments}

 \bibliographystyle{elsarticle-num-names}

\bibliography{bib/dual-numbers/cite}

\begin{thebibliography}{16}
\expandafter\ifx\csname natexlab\endcsname\relax\def\natexlab#1{#1}\fi
\providecommand{\url}[1]{\texttt{#1}}
\providecommand{\href}[2]{#2}
\providecommand{\path}[1]{#1}
\providecommand{\DOIprefix}{}
\providecommand{\ArXivprefix}{arXiv:}
\providecommand{\URLprefix}{URL: }
\providecommand{\Pubmedprefix}{pmid:}
\providecommand{\doi}[1]{\href{http://dx.doi.org/#1}{\path{#1}}}
\providecommand{\Pubmed}[1]{\href{pmid:#1}{\path{#1}}}
\providecommand{\bibinfo}[2]{#2}
\ifx\xfnm\relax \def\xfnm[#1]{\unskip,\space#1}\fi
\bibitem[{Yaglom et~al.(1964)Yaglom, Rozenfel'd, and
  Yasinskaya}]{yaglom:1964:projective-metrics::en}
\bibinfo{author}{I.~M. Yaglom}, \bibinfo{author}{B.~A. Rozenfel'd},
  \bibinfo{author}{E.~U. Yasinskaya},
\newblock \bibinfo{title}{{Projective Metrics}},
\newblock \bibinfo{journal}{Russian Mathematical Surveys} \bibinfo{volume}{19}
  (\bibinfo{year}{1964}) \bibinfo{pages}{49--107}.
  \DOIprefix\doi{10.1070/RM1964v019n05ABEH001159}.
\bibitem[{Yaglom(1968)}]{yaglom:complex-numbers-geometry::en}
\bibinfo{author}{I.~M. Yaglom}, \bibinfo{title}{{Complex Numbers in Geometry}},
  \bibinfo{publisher}{Academic Press}, \bibinfo{year}{1968}.
\bibitem[{Rozenfel'd and Yaglom(1951)}]{yaglom:1951:geometry-algebra::en}
\bibinfo{author}{B.~A. Rozenfel'd}, \bibinfo{author}{I.~M. Yaglom},
\newblock \bibinfo{title}{{On the geometries of the simplest algebras}},
\newblock \bibinfo{journal}{Mat. Sbornik N. S.} \bibinfo{volume}{28(70)}
  (\bibinfo{year}{1951}) \bibinfo{pages}{205--216}.
\bibitem[{Kulyabov et~al.(2020)Kulyabov, Korolkova, and
  Gevorkyan}]{kulyabov:2020:jopcs:hyperbolic-numbers}
\bibinfo{author}{D.~S. Kulyabov}, \bibinfo{author}{A.~V. Korolkova},
  \bibinfo{author}{M.~N. Gevorkyan},
\newblock \bibinfo{title}{{Hyperbolic numbers as Einstein numbers}},
\newblock \bibinfo{journal}{Journal of Physics: Conference Series}
  \bibinfo{volume}{1557} (\bibinfo{year}{2020}) \bibinfo{pages}{012027.1--5}.
  \DOIprefix\doi{10.1088/1742-6596/1557/1/012027}.
\bibitem[{Cayley(1859)}]{cayley:1859:sixth-memoir-upon-quantics}
\bibinfo{author}{A.~Cayley},
\newblock \bibinfo{title}{{IV. A sixth memoir upon quantics}},
\newblock \bibinfo{journal}{Philosophical Transactions of the Royal Society of
  London} \bibinfo{volume}{149} (\bibinfo{year}{1859}) \bibinfo{pages}{61--90}.
  \DOIprefix\doi{10.1098/rstl.1859.0004}.
\bibitem[{Klein(1985)}]{klein:1985:nicht-euklidische}
\bibinfo{author}{F.~Klein},
\newblock \bibinfo{title}{{Ueber die sogenannte Nicht-Euklidische Geometrie}},
\newblock in: \bibinfo{booktitle}{Gau{\ss} und die Anf{\"{a}}nge der
  nicht-euklidischen Geometrie}, volume~\bibinfo{volume}{4} of
  \textit{\bibinfo{series}{Teubner-Archiv zur Mathematik}},
  \bibinfo{publisher}{Springer-Verlag Wien}, \bibinfo{address}{Wien},
  \bibinfo{year}{1985}, pp. \bibinfo{pages}{224--238}.
  \DOIprefix\doi{10.1007/978-3-7091-9511-6_5}.
\bibitem[{L_J(2020{\natexlab{a}})}]{L_Julia:2020:ForwardDiff}
\bibinfo{title}{{Forward Mode Automatic Differentiation for Julia}},
  \bibinfo{year}{2020}{\natexlab{a}}. \URLprefix
  \url{https://github.com/JuliaDiff/DualNumbers.jl}.
\bibitem[{L_J(2020{\natexlab{b}})}]{L_Julia:2020:site}
\bibinfo{title}{{The Julia Language}}, \bibinfo{year}{2020}{\natexlab{b}}.
  \URLprefix \url{https://julialang.org/}.
\bibitem[{Bezanson et~al.(2018)Bezanson, Chen, Chung, Karpinski, Shah, Vitek,
  and Zoubritzky}]{bezanson:2018:julia-dinamism-by-design}
\bibinfo{author}{J.~Bezanson}, \bibinfo{author}{J.~Chen},
  \bibinfo{author}{B.~Chung}, \bibinfo{author}{S.~Karpinski},
  \bibinfo{author}{V.~B. Shah}, \bibinfo{author}{J.~Vitek},
  \bibinfo{author}{L.~Zoubritzky},
\newblock \bibinfo{title}{{Julia: dynamism and performance reconciled by
  design}},
\newblock \bibinfo{journal}{Proceedings of the ACM on Programming Languages}
  \bibinfo{volume}{2} (\bibinfo{year}{2018}) \bibinfo{pages}{1--23}.
  \DOIprefix\doi{10.1145/3276490}.
\bibitem[{{Zappa Nardelli} et~al.(2018){Zappa Nardelli}, Belyakova, Pelenitsyn,
  Chung, Bezanson, and Vitek}]{zappa:2018:julia-subtyping}
\bibinfo{author}{F.~{Zappa Nardelli}}, \bibinfo{author}{J.~Belyakova},
  \bibinfo{author}{A.~Pelenitsyn}, \bibinfo{author}{B.~Chung},
  \bibinfo{author}{J.~Bezanson}, \bibinfo{author}{J.~Vitek},
\newblock \bibinfo{title}{{Julia subtyping: a rational reconstruction}},
\newblock \bibinfo{journal}{Proceedings of the ACM on Programming Languages}
  \bibinfo{volume}{2} (\bibinfo{year}{2018}) \bibinfo{pages}{1--27}.
  \DOIprefix\doi{10.1145/3276483}.
\bibitem[{Bezanson et~al.(2017)Bezanson, Edelman, Karpinski, and
  Shah}]{bezanson:2017:julia-numeric-computing}
\bibinfo{author}{J.~Bezanson}, \bibinfo{author}{A.~Edelman},
  \bibinfo{author}{S.~Karpinski}, \bibinfo{author}{V.~B. Shah},
\newblock \bibinfo{title}{{Julia: A fresh approach to numerical computing}},
\newblock \bibinfo{journal}{SIAM Review} \bibinfo{volume}{59}
  (\bibinfo{year}{2017}) \bibinfo{pages}{65--98}.
  \DOIprefix\doi{10.1137/141000671}. \href{http://arxiv.org/abs/1411.1607}{{\tt
  arXiv:1411.1607}}.
\bibitem[{Bezanson et~al.(2012)Bezanson, Karpinski, Shah, and
  Edelman}]{bezanson:2012:julia-dinamic-language}
\bibinfo{author}{J.~Bezanson}, \bibinfo{author}{S.~Karpinski},
  \bibinfo{author}{V.~B. Shah}, \bibinfo{author}{A.~Edelman},
\newblock \bibinfo{title}{{Julia: A Fast Dynamic Language for Technical
  Computing}}  (\bibinfo{year}{2012}) \bibinfo{pages}{1--27}.
  \href{http://arxiv.org/abs/1209.5145}{{\tt arXiv:1209.5145}}.
\bibitem[{L_J(2020{\natexlab{a}})}]{L_Julia:2020:DualNumbers}
\bibinfo{title}{{Julia package for representing dual numbers and for performing
  dual algebra}}, \bibinfo{year}{2020}{\natexlab{a}}. \URLprefix
  \url{https://github.com/JuliaDiff/ForwardDiff.jl}.
\bibitem[{L_J(2020{\natexlab{b}})}]{L_Julia:repo:2020}
\bibinfo{title}{{Official Julia language GitHub repository}},
  \bibinfo{year}{2020}{\natexlab{b}}. \URLprefix
  \url{https://github.com/JuliaLang/julia}.
\bibitem[{Hamilton(1866)}]{hamilton:1866:elements-quaternions}
\bibinfo{author}{W.~R. Hamilton}, \bibinfo{title}{{Elements of Quaternions}},
  \bibinfo{publisher}{Cambridge University Press},
  \bibinfo{address}{Cambridge}, \bibinfo{year}{1866}.
  \DOIprefix\doi{10.1017/CBO9780511707162}.
\bibitem[{Yefremov(2005)}]{yefremov:2005:quaternions-biquaternions}
\bibinfo{author}{A.~P. Yefremov}, \bibinfo{title}{{Quaternions and
  Biquaternions: Algebra, Geometry and Physical Theories}},
  \bibinfo{year}{2005}. \href{http://arxiv.org/abs/0501055}{{\tt
  arXiv:0501055}}.

\end{thebibliography}


\begin{thebibliography}{10}
\def\selectlanguageifdefined#1{
\expandafter\ifx\csname date#1\endcsname\relax
\else\selectlanguage{#1}\fi}
\providecommand*{\href}[2]{{\small #2}}
\providecommand*{\url}[1]{{\small #1}}
\providecommand*{\BibUrl}[1]{\url{#1}}
\providecommand{\BibAnnote}[1]{}
\providecommand*{\BibEmph}[1]{#1}
\ProvideTextCommandDefault{\cyrdash}{\iflanguage{russian}{\hbox
  to.8em{--\hss--}}{\textemdash}}
\providecommand*{\BibDash}{\ifdim\lastskip>0pt\unskip\nobreak\hskip.2em plus
  0.1em\fi
\cyrdash\hskip.2em plus 0.1em\ignorespaces}
\renewcommand{\newblock}{\ignorespaces}

\bibitem{yaglom:1964:projective-metrics::ru}
\selectlanguageifdefined{russian}
\BibEmph{Яглом~И.~М., Розенфельд~Б.~А., Ясинская~Е.~У.} {Проективные
  метрики}~// \BibEmph{Успехи математических наук}. \BibDash
\newblock 1964. \BibDash
\newblock Т.~19, {№} 5 (119). \BibDash
\newblock {С.}~51--113.

\bibitem{yaglom:complex-numbers-geometry::ru}
\selectlanguageifdefined{russian}
\BibEmph{Яглом~И.~М.} {Комплексные числа и их применение в геометрии}~//
  Математика, ее преподавание, приложения и история. \BibDash
\newblock 1961. \BibDash
\newblock Т.~6 {из} \BibEmph{Математическое просвещение, сер. 2}. \BibDash
\newblock {С.}~61--106.

\bibitem{yaglom:1951:geometry-algebra::ru}
\selectlanguageifdefined{russian}
\BibEmph{Розенфельд~Б.~А., Яглом~И.~М.} {О геометриях простейших алгебр}~//
  \BibEmph{Математический сборник}. \BibDash
\newblock 1951. \BibDash
\newblock Т. 28(70), {№}~1. \BibDash
\newblock {С.}~205--216.

\bibitem{kulyabov:2020:jopcs:hyperbolic-numbers}
\selectlanguageifdefined{english}
\BibEmph{Kulyabov~D.~S., Korolkova~A.~V., Gevorkyan~M.~N.} {Hyperbolic numbers
  as Einstein numbers}~//
  \href{http://dx.doi.org/10.1088/1742-6596/1557/1/012027}{\BibEmph{Journal of
  Physics: Conference Series}}. \BibDash
\newblock 2020. \BibDash 5. \BibDash
\newblock Vol. 1557. \BibDash
\newblock P.~012027.1--5.

\bibitem{cayley:1859:sixth-memoir-upon-quantics}
\selectlanguageifdefined{english}
\BibEmph{Cayley~A.} {IV. A sixth memoir upon quantics}~//
  \href{http://dx.doi.org/10.1098/rstl.1859.0004}{\BibEmph{Philosophical
  Transactions of the Royal Society of London}}. \BibDash
\newblock 1859. \BibDash 1. \BibDash
\newblock Vol. 149. \BibDash
\newblock P.~61--90.

\bibitem{klein:1985:nicht-euklidische}
\selectlanguageifdefined{german}
\BibEmph{Klein~F.} \href{http://dx.doi.org/10.1007/978-3-7091-9511-6_5}{{Ueber
  die sogenannte Nicht-Euklidische Geometrie}}~// Gau{\ss} und die
  Anf{\"{a}}nge der nicht-euklidischen Geometrie. \BibDash
\newblock Wien~: Springer-Verlag Wien, 1985. \BibDash
\newblock {Bd.}~4 von \BibEmph{Teubner-Archiv zur Mathematik}. \BibDash
\newblock S.~224--238.

\bibitem{L_Julia:2020:ForwardDiff}
\selectlanguageifdefined{english}
{Forward Mode Automatic Differentiation for Julia}. \BibDash
\newblock 2020. \BibDash
\newblock Access mode: \BibUrl{https://github.com/JuliaDiff/DualNumbers.jl}.

\bibitem{L_Julia:2020:site}
\selectlanguageifdefined{english}
{The Julia Language}. \BibDash
\newblock 2020. \BibDash
\newblock Access mode: \BibUrl{https://julialang.org/}.

\bibitem{bezanson:2018:julia-dinamism-by-design}
\selectlanguageifdefined{english}
\BibEmph{Bezanson~J., Chen~J., Chung~B., Karpinski~S., Shah~V.~B., Vitek~J.,
  Zoubritzky~L.} {Julia: dynamism and performance reconciled by design}~//
  \href{http://dx.doi.org/10.1145/3276490}{\BibEmph{Proceedings of the ACM on
  Programming Languages}}. \BibDash
\newblock 2018. \BibDash 10. \BibDash
\newblock Vol.~2, no. OOPSLA. \BibDash
\newblock P.~1--23.

\bibitem{zappa:2018:julia-subtyping}
\selectlanguageifdefined{english}
\BibEmph{{Zappa Nardelli}~F., Belyakova~J., Pelenitsyn~A., Chung~B.,
  Bezanson~J., Vitek~J.} {Julia subtyping: a rational reconstruction}~//
  \href{http://dx.doi.org/10.1145/3276483}{\BibEmph{Proceedings of the ACM on
  Programming Languages}}. \BibDash
\newblock 2018. \BibDash 10. \BibDash
\newblock Vol.~2, no. OOPSLA. \BibDash
\newblock P.~1--27.

\bibitem{bezanson:2017:julia-numeric-computing}
\selectlanguageifdefined{english}
\BibEmph{Bezanson~J., Edelman~A., Karpinski~S., Shah~V.~B.} {Julia: A fresh
  approach to numerical computing}~//
  \href{http://dx.doi.org/10.1137/141000671}{\BibEmph{SIAM Review}}. \BibDash
\newblock 2017. \BibDash 1. \BibDash
\newblock Vol.~59, no.~1. \BibDash
\newblock P.~65--98. \BibDash
\newblock arXiv~: 1411.1607.

\bibitem{bezanson:2012:julia-dinamic-language}
\selectlanguageifdefined{english}
\BibEmph{Bezanson~J., Karpinski~S., Shah~V.~B., Edelman~A.} {Julia: A Fast
  Dynamic Language for Technical Computing}. \BibDash
\newblock 2012. \BibDash
\newblock P.~1--27. \BibDash
\newblock arXiv~: 1209.5145.

\bibitem{L_Julia:2020:DualNumbers}
\selectlanguageifdefined{english}
{Julia package for representing dual numbers and for performing dual algebra}.
  \BibDash
\newblock 2020. \BibDash
\newblock Access mode: \BibUrl{https://github.com/JuliaDiff/ForwardDiff.jl}.

\bibitem{L_Julia:repo:2020}
\selectlanguageifdefined{english}
{Official Julia language GitHub repository}. \BibDash
\newblock 2020. \BibDash
\newblock Access mode: \BibUrl{https://github.com/JuliaLang/julia}.

\bibitem{hamilton:1866:elements-quaternions}
\selectlanguageifdefined{english}
\BibEmph{Hamilton~W.~R.}
  \href{http://dx.doi.org/10.1017/CBO9780511707162}{{Elements of
  Quaternions}}~/ Ed.\ by\ W.~E.~Hamilton. \BibDash
\newblock Cambridge~: Cambridge University Press, 1866. \BibDash
\newblock 762~p. \BibDash
\newblock
  ISBN:~\href{http://isbndb.com/search-all.html?kw=9780511707162}{9780511707162}.

\bibitem{yefremov:2005:quaternions-biquaternions}
\selectlanguageifdefined{english}
\BibEmph{Yefremov~A.~P.} {Quaternions and Biquaternions: Algebra, Geometry and
  Physical Theories}. \BibDash
\newblock 2005. \BibDash 1. \BibDash
\newblock arXiv~: math-ph/0501055.

\end{thebibliography}

\end{document}


\graphicspath{{image/dual-numbers/ru/}{image/dual-numbers/}{image/}}

\title{Подходы к реализации обобщённых комплексных чисел на языке Julia}

\author{М. Н. Геворкян}
\email{gevorkyan-mn@rudn.ru}
\affiliation{Кафедра прикладной информатики и теории вероятностей,\\
  Российский университет дружбы народов,\\
  117198, Москва, ул. Миклухо-Маклая, д. 6}

\author{А. В. Королькова}
\email{korolkova-av@rudn.ru}
\affiliation{Кафедра прикладной информатики и теории вероятностей,\\
  Российский университет дружбы народов,\\
  117198, Москва, ул. Миклухо-Маклая, д. 6}

\author{Д. С. Кулябов}
\email{kulyabov-ds@rudn.ru}
\affiliation{Кафедра прикладной информатики и теории вероятностей,\\
  Российский университет дружбы народов,\\
  117198, Москва, ул. Миклухо-Маклая, д. 6}
\affiliation{Лаборатория информационных технологий,\\
  Объединённый институт ядерных исследований,\\
  ул. Жолио-Кюри 6, Дубна, Московская область, Россия, 141980}

\begin{abstract}

  В задачах математической физики для исследования структур
  пространств с применением моделей Кэли--Клейна в теоретических
  расчётах требуется использование обобщённых комплексных чисел.  В
  случае проведения вычислительных экспериментов для такого рода задач
  необходима их качественная реализация в языке
  программирования. Предлагаемые малочисленные реализации обобщённых
  комплексных чисел в современных языках программирования имеют ряд
  недостатков. В данной статье мы предлагаем в качестве языка
  реализации обобщённых комплексных чисел использовать язык Julia, не
  в последнюю очередь из-за поддержки им механизма множественной
  диспетчеризации.

  В работе демонстрируется подход к реализации одного из типов
  обобщённых комплексных чисел, а именно дуальных чисел. Особый упор
  мы делаем на описание использования механизма множественной
  диспетчеризации для реализации нового числового типа. Полученную
  реализацию дуальных чисел можно рассматривать как прототип для
  полного программного модуля поддержки обобщённых комплексных чисел.

\end{abstract}

  \keywords{%
    комплексные числа, параболические комплексные числа,
    дуальные числа,
    множественная диспетчеризация,
    Julia}

\maketitle

\section{Введение}
\label{sec:intro}

  Существует подход, который позволяет обобщить обычные комплексные
  числа и получить три различных класса обобщённых комплексных
  чисел~\cite{yaglom:1964:projective-metrics::ru,yaglom:complex-numbers-geometry::ru,yaglom:1951:geometry-algebra::ru,kulyabov:2020:jopcs:hyperbolic-numbers}.
  В этом подходе для определения трёх видов комплексных чисел задаётся
  квадратное уравнение
\begin{equation}
  \label{eq:x^2+px+q}
  z^2 + p z +q =0.
\end{equation}
с детерминантом
\begin{equation}
  \label{eq:determinant}
  \Delta = p^2 -4 p q.
\end{equation}
В зависимости от знака детерминанта получаются следующие 
комплексные числа:
\begin{itemize}
\item $\Delta < 0$~--- эллиптические  (обычные комплексные числа);
\item $\Delta = 0$~--- параболические  (дуальные комплексные числа);
\item $\Delta > 0$~--- гиперболические  (двойные комплексные  (расщеплённые)
  числа).
\end{itemize}

Первоначально эти системы комплексных чисел вводились для описания
моделей
Кэли--Клейна~\cite{cayley:1859:sixth-memoir-upon-quantics,klein:1985:nicht-euklidische}.
Однако они имеют и другие применения. В частности, дуальные
комплексные числа могут использоваться в задачах автоматического
дифференцирования~\cite{L_Julia:2020:ForwardDiff}.

Интерес представляет чистая реализация обобщённых комплексных чисел,
не привязанная к частным применениям. Существующие реализации
обобщённых комплексных чисел в рамках пакетов языка Julia --- не
удовлетворительны.  В данной статье мы предлагаем свой подход по
реализации дуальных комплексных чисел в Julia.

\subsection{Структура статьи}
\label{sec:structure}
В разделе~\ref{sec:dual-complex} приводится достаточно подробное
описание дуальных комплексных чисел. Описание конструктивное,
затрагивающее только операции с дуальными комплексными числами. Мы
постарались избегать излишних математических абстракций.  В
разделе~\ref{sec:julia-features} даётся общее понятие о механизме
множественной диспетчеризации в Julia. На использовании этого
механизма и базируется предлагаемая нами реализация дуальных
комплексных чисел.  В разделе~\ref{sec:dual-complex-julia} собственно
и описывается наша реализация дуальных комплексных чисел на Julia.

\subsection{Обозначения и соглашения}
\label{sec:notation}

  Для обозначения дуальной комплексной единицы мы будем использовать
  символ $\iiz$.

\section{Дуальные числа}
\label{sec:dual-complex}

Дуальное число $z$ алгебраически определяется следующим образом:
\begin{equation}
  z = a + \iiz b,\; a,b\in\setR,\; \iiz^2 = 0, \iiz \neq 0.
\end{equation}
Величину $a$ будем называть действительной частью, а $b$ ---
мнимой. Так как в рамках данной статьи не будет встречаться обычных
комплексных чисел, то путаницы в терминологии не возникнет.

\subsection{Алгебраическая форма}
Алгебраические свойства для двух чисел $z_1 = a_1 + \iiz b_1$ и
$z_2 = a_2 + \iiz b_2$.
\begin{description}
  \item[Сложение] $z_1 + z_2 = (a_1 + a_2) + \iiz (b_1 + b_2)$.
  \item[Вычитание] $z_1 - z_2 = (a_1 - a_2) + \iiz (b_1 - b_2)$.
  \item[Умножение] $z_1 \cdot z_2 = (a_1 + \iiz b_1) \cdot (a_2 + \iiz b_2) = a_1 a_2 + \iiz (a_1 b_2 + b_1 a_2)$.
  \item[Сопряжённое число] $\bar{z} = a - \iiz b$,
    $z \bar{z} = (a + \iiz b) \cdot (a - \iiz b) = a^2$.
  \item[Абсолютное значение числа] $|z| = a$.

    Модуль дуального числа
    может быть отрицательным, кроме того его можно вычислить по
    следующей формуле:
    \begin{equation}
      |z| = \dfrac12 (z + \bar{z}) = \dfrac12 (a + \iiz
      b + a - \iiz b) = a.
    \end{equation}
  \item[Деление] Деление двух дуальных чисел $z_1$ и $z_2$ определено
    для всех $z_2$ таких, что $|z_2| \neq 0$:
  \begin{equation}
    \dfrac{z_1}{z_2} = \dfrac{a_1 + \iiz b_1}{a_2 + \iiz b_2} =
    \dfrac{(a_1 + \iiz b_1)(a_2 - \iiz b_2)}{(a_2 + \iiz b_2)(a_2 -
      \iiz b_2)} = \dfrac{a_1}{a_2} + \iiz \dfrac{b_1 a_2 - b_2
      a_1}{a^2_2}.    
  \end{equation}    
\end{description}

Из вышеперечисленных свойств видно, что ни в одном случае мнимые части чисел не вносят вклад в действительную часть результата.

Рассмотрим дуальное число вида $c\iiz$, где $c \in \setR$. Это дуальное число с нулевым абсолютным значением. Такие числа обладают следующим свойством:
\begin{equation}
  (c\iiz) \cdot (d\iiz) = cd\iiz^2 = 0,  
\end{equation}
из которого следует, что для любого числа $c\iiz$ существует число $d\iiz$, такое что произведение этих чисел равняется $0$. Такие числа называются \emph{делителями нуля}.

\subsection{Тригонометрическая форма}

Дуальное число $z$, такое что $|z| \neq 0$, можно записать в следующем виде:
\begin{equation}
  a + \iiz b = a \Big( 1 + \dfrac{b}{a}\iiz \Big) = r(1 + \varphi \iiz),  
\end{equation}
где величины $r = |z| = a$ и $\iiz = Arg z = b/a$ называются \emph{модулем} и \emph{аргументом} дуального числа $z$ соответственно.
 
Такая форма записи дуального числа является некоторым аналогом тригонометрической формы обычного комплексного числа и далее будем её называть \emph{тригонометрической формой} дуального числа.

Для сопряжённого числа $\bar{z}$ тригонометрическая форма имеет вид:
\begin{equation}
  \bar{z} = r(1 - \varphi \iiz),
\end{equation}
При этом $|\bar{z}| = |z|$ --- модуль, а $Arg\bar{z} = -b/a$ --- аргумент дуального комплексного числа в тригонометрической форме.

Аналогия с тригонометрической формой комплексного числа продолжается далее при рассмотрении умножения и деления. При умножении двух дуальных чисел $z_1$ и $z_2$ получим
\begin{equation}
  z = r_1(1 + \varphi_1 \iiz) \cdot r_2(1 + \varphi_2 \iiz) = r_1 r_2 \Big(1 + (\varphi_1 + \varphi_2)\iiz\Big),
\end{equation}
то есть при умножении аргументы складываются, а модули умножаются.

В случае же деления, получим
\begin{equation}
  \dfrac{z_1}{z_2} = \dfrac{r_1(1+\varphi_1 \iiz_1)}{r_2(1 + \varphi_2\iiz)} = \dfrac{r_1 (1 + \varphi_1 \iiz) r_2 (1 - \varphi_2\iiz)}{r_2(1 + \varphi_2\iiz)r_2(1 - \varphi_2\iiz)} = \dfrac{r_1(1 + (\varphi_1 - \varphi_2)\iiz)}{r_2 (1 - \varphi_2 \iiz + \varphi_2 \iiz)} = \dfrac{r_1}{r_2}(1 + (\varphi_1 - \varphi_2)\iiz),
\end{equation}
то есть при делении аргументы вычитаются, а модули делятся.

Обратите внимание, что в случае деления число $z_2$ должно иметь ненулевой модуль $|z_2| \neq 0$.

В тригонометрической форме особенно просто выглядит операция возведения в натуральную степень $n$:
\begin{equation}
  z^n = \big(r (1 + \varphi \iiz)\big)^n = \underbrace{r\cdot r \cdot \ldots \cdot r}_{n}\big(1 + \underbrace{(\varphi + \varphi + \ldots + \varphi)}_{n}\big) = r^n (1 + n\varphi\iiz) = a^n + \iiz n a^{n-1}b.
\end{equation}

Для нахождения
$\sqrt[n]{z} = \sqrt[n]{r(1 + \varphi\iiz)}$, $n \in \mathbb{N}$, предположим, что
$\sqrt[n]{z} = z_1$, тогда по определению $z^n_1 = z$, следовательно
\begin{equation}
  r_1^n(1 + n\varphi_1\iiz) = z = r(1 + \varphi\iiz),
\end{equation}
что даёт
\begin{equation}
  r_1 = \sqrt[n]{r}, \varphi_1 = \dfrac{\varphi}{n}, \sqrt[n]{z} = \sqrt[n]{r}\left(1 + \dfrac{\varphi}{n}\iiz\right).
\end{equation}

Следует заметить, что при нечётном $n$ корень $\sqrt[n]{z}$ существует всегда, а при чётном $n$ --- только для числа $z$ с неотрицательным модулем $|z| = r \neq 0$. В алгебраической форме формула для извлечения корня имеет вид:
\begin{equation}
  \sqrt[n]{a + b\iiz} = \sqrt[n]{a} + \dfrac{b a^{\frac{1-n}{n}}}{n}\iiz.
\end{equation}

\subsection{Матричная форма}
Все алгебраические операции над дуальными числами можно свести к матричным операциям, если положить
\begin{equation}
  \iiz \leftrightarrow 
  \begin{pmatrix}
    0 & 1 \\
    0 & 0
  \end{pmatrix}
  \;\;
  z = a + b\iiz \leftrightarrow
  \begin{pmatrix}
    a & b \\
    0 & a
  \end{pmatrix}.
\end{equation}

Тогда, например:
\begin{equation}
  z_1 \cdot z_2 \leftrightarrow 
  \begin{pmatrix}
    a_1 & b_1 \\
    0 & a_1
  \end{pmatrix}
  \begin{pmatrix}
    a_2 & b_2 \\
    0 & a_2
  \end{pmatrix}
  =
  \begin{pmatrix}
    a_1a_2 & a_1b_2 + a_2 b_1 \\
    0 & a_1a_2
  \end{pmatrix}
  \leftrightarrow
  a_1a_2 + (a_1b_2 + a_2 b_1)\iiz.
\end{equation}

Теоретически с вычислительной точки зрения сводить действия над дуальными числами к матричным операциям может быть оправданно в случае, если язык программирования поддерживает векторизацию операций. Однако вряд ли на практике эффект будет значителен.

\subsection{Разложение в ряд Тейлора}

Используя свойство
$\iiz^2 = \iiz^3 = \ldots = \iiz^n = 0$ $\forall n \in \mathbb{N}$, получим:
\begin{equation}
  \exp(b\iiz) = \sum\limits^{\infty}_{n=0}\dfrac{(b\iiz)^n}{n!} = 1 + b\iiz + \dfrac{b^2\iiz^2}{2!} + \ldots = 1 + b\iiz,
\end{equation}
\begin{equation}
  \exp(a + b\iiz) = e^ae^{b\iiz} = e^a(1 + b\iiz).
\end{equation}

Более общая формула выводится из ряда Тейлора для функции $f(z)$ в точке $a$:
\begin{equation}
  f(a + \iiz b) = \sum\limits^{\infty}_{n=0}\dfrac{f^{(n)}(a)(a + \iiz b - a)^n}{n!} = \sum\limits^{\infty}_{n=0}\dfrac{f^{(n)}(a)\iiz^nb^n}{n!} = f(a) + f'(a)b\iiz + \dfrac{f^{''}(a)\iiz^2b^2}{2!} + \ldots = f(a) + f'(a)b\iiz.
\end{equation}
В результате получаем крайне важную формулу:
\begin{equation}
  f(a + \iiz b) = f(a) + f'(a)b\iiz,
\end{equation}
которая даёт способ вычисления значений функций от дуального числа, если известно значение производной $f'(a)$. С другой стороны, эта же формула позволяет вычислить значение производной в точке $a$.

\subsection{Элементарные функции от дуальных чисел}

Формула $f(a + \iiz b) = f(a) + f'(a)b\iiz$ позволяет распространить элементарные функции на множество дуальных чисел, так как в правой части формулы находятся лишь значения функции $f$ от действительного числа $a$.

Приведём для иллюстрации небольшую сводку основных элементарных функций.

\begin{center}
  {\renewcommand{\arraystretch}{1.2}%
  \begin{tabular}{l|l}
    Trigonometric function & Inverse trigonometric functions \\ \hline
    $\sin(a+\iiz b) = \sin{a} + b \iiz\cos{a}$ & $\arcsin(a + \iiz b) = \arcsin{a} + b\iiz / \sqrt{1 - a^2}$\\
    $\cos(a+\iiz b) = \cos{a} - b \iiz\sin{a}$ & $\arccos(a + \iiz b) = \arccos{a} - b\iiz / \sqrt{1 - a^2}$\\
    $\tg(a+\iiz b) = \tg{a} + b \iiz / \cos^2{a}$ & $\arctg(a + \iiz b) = \arctg{a} + b\iiz / (1 + a^2)$\\
    $\ctg(a+\iiz b) = \ctg{a} - b \iiz / \sin^2{a}$ & $\arcctg(a + \iiz b) = \arctg{a} - b\iiz / (1 + a^2)$
  \end{tabular}}
\end{center}

\begin{center}
  {\renewcommand{\arraystretch}{1.2}%
  \begin{tabular}{l|l}
    Power functions & Logarithmic functions and exponent\\ \hline
    $(a + \iiz b)^n = a^n + n a^{n-1} b \iiz$ & $\exp(a + \iiz b) = \exp{a} + b\iiz \exp{a}$\\
    $\sqrt[n]{a + \iiz} = \sqrt[n]{a}\Big( 1 + \dfrac{b\iiz}{n a} \Big)$ & $\log_{c}(a + \iiz b) = \log_c{a} + b\iiz / a \ln{a}$
  \end{tabular}}
\end{center}

\subsection{Вычисление первой производной от вещественной функции с помощью дуальных чисел}

Формулу $f(a + \iiz b) = f(a) + f'(a)b\iiz$ можно применять и для вычисления значения первой производной вещественной функции $f(x)$ в точке $a$, если известно значении функции $f(a + \iiz b)$. С точки зрения аналитических вычислений такой способ нахождения производной не имеет смысла, так как сама величина $f(a + \iiz b)$ получается аналитически из известной формулы для первой производной $f'(x)$. Однако она даёт численный метод для получения значения первой производной с помощью компьютерных вычислений без какой-либо дополнительной погрешности.  Такой способ нахождения производной получил название \emph{автоматического дифференцирования} с использованием дуальных чисел.

Рассмотрим вначале простой пример автоматического дифференцирования, а далее обсудим общий принцип реализации для языков программирования.

В качестве примера найдём производную от функции $f(x) = x\sin{x}$:
\begin{equation}
  f(a + \iiz b) = (a + \iiz b) \sin(a + \iiz b).
\end{equation}

Зная значение $\sin(a + \iiz b) = \sin{a} + b\iiz\cos{a}$, легко вычислить $f(a + \iiz b)$:
\begin{equation}
  f(a + \iiz b) = (a + \iiz)(\sin{a} + b\iiz\cos{a}) = a\sin{a} + (\sin{a} + a\cos{a})b\iiz.
\end{equation}

Сравнивая с общей формулой $f(a + \iiz b) = f(a) + f'(a)b\iiz$, получаем значение $f'(a) = \sin{a} + b\iiz\cos{a}$.

В общем случае для нахождения $f'(a)$ достаточно найти значение $f'(a + \iiz b)$, взять мнимую часть этого дуального числа и поделить на $b$. Кроме того, так как выбор числа $b$ произволен, то можно выбрать его равным $1$ и избавиться от необходимости деления на $b$.

Для применения автоматического дифференцирования с помощью дуальных чисел в языке программирования должен быть определён тип дуальных чисел, а также арифметические операции и элементарные функции для данного типа. Особенно просто это осуществляется для объектно-ориентированных языков и языков, поддерживающих перегрузку функций или множественную диспетчеризацию.  Также следует иметь в виду, что автоматическое дифференцирование с помощью дуальных чисел ограничено производными первого порядка.

\section{Динамическая диспетчеризация в языке Julia}

\label{sec:julia-features}

  Язык Julia~\cite{L_Julia:2020:site} появился сравнительно недавно,
  однако успел приобрести популярность как язык для научных
  вычислений. Будем предполагать, что читатели уже знакомы с этим
  языком, и кратко остановимся лишь на концепции \emph{множественной
    диспетчеризации}~\cite{bezanson:2018:julia-dinamism-by-design,zappa:2018:julia-subtyping,bezanson:2017:julia-numeric-computing,bezanson:2012:julia-dinamic-language},
  которая лежит в основе языка, её понимание существенно для
  дальнейшего изложения.

  \emph{Динамическая диспетчеризация} (dynamic dispatch) --- это
  механизм, который позволяет выбрать конкретную реализацию
  полиморфной функции или оператора из множества и вызвать в
  конкретном случае.

  \emph{Множественная диспетчеризация} основывается на динамической
  диспетчеризации. В этом случае выбор реализации полиморфной функции
  делается исходя из типа, количества и порядка следования аргументов
  функции. Таким образом реализуется полиморфизм времени выполнения
  (runtime polymorphic dispatch). Кроме термина <<множественная
  диспетчеризация>> также употребляется термин \emph{мультиметод}.

  Механизм множественной диспетчеризации похож на механизм перегрузки
  функций и операторов, реализованный, например, в языке
  C++. Перегрузка функций, однако, осуществляется исключительно на
  стадии компиляции, тогда как множественная диспетчеризация должна
  работать также и на стадии выполнения программы (полиморфизм времени
  выполнения).

  Julia поддерживает перегрузку функций на стадии компиляции в случае,
  если все типы данных, используемые в функции, выводимы на стадии
  компиляции (type stable functions). При этом JIT-компилятор создаёт
  эффективные реализации для каждой комбинации типов аргументов.

  Если же типы данных невыводимы на стадии компиляции (type unstable),
  то включается механизм динамической диспетчеризации. Компилятор не
  сможет создать специализированную версию, а создаст универсальную,
  которая работает медленно.

  Такой подход позволяет совмещать быстродействие компилируемого языка
  со строгой статической типизацией с гибкостью интерпретируемого
  языка с динамической типизацией.

\section{Реализация дуальных чисел на языке Julia}
\label{sec:dual-complex-julia}

\subsection{Существующие реализации дуальных чисел на Julia}

Авторам известны по меньшей мере две реализации дуальных чисел на языке Julia:
\begin{itemize}
\item тип \verb|Dual| в модуле для автоматического дифференцирования   Forward\-Diff~\cite{L_Julia:2020:ForwardDiff};
\item отдельный модуль Dual\-Numbers~\cite{L_Julia:2020:DualNumbers},   развитие которого заморожено в пользу ForwardDiff.
\end{itemize}

В данном разделе мы опишем предлагаемую нами реализацию дуальных чисел на языке Julia. Она основана как на встроенном типе обычных комплексных чисел \verb|Complex|~\cite{L_Julia:repo:2020}, так и на модуле Dual\-Numbers. Предлагаемая в данной работе реализация служит исключительно в качестве учебного примера создания пользовательского типа данных для языка Julia.

Во главу угла мы поставили ясность изложения, поэтому многие вычислительные оптимизации были умышленно опущены в пользу большей ясности. Так, например, в модуле Dual\-Numbers дуальное число $z = a + \iiz b$ может иметь комплексные компоненты $a$ и $b$. Также в Dual\-Numbers для реализации элементарных функций от дуальных чисел $f(z)$ используется сторонний модуль, где определены символьные правила дифференцирования для функций от действительного переменного. Это позволяет автоматизировать определение функции $f(z)$ по формуле $f(a + \iiz b) = f(a) + f'(a)b\iiz$, так как не нужно явно выписывать производные $f'(a)$.

В нашей реализации $a$ и $b$ --- исключительно вещественные, а элементарные функции определены явно.

\subsection{Объявление структуры данных}

\begin{figure}[!h]
  \centering
  \includegraphics[width=\linewidth]{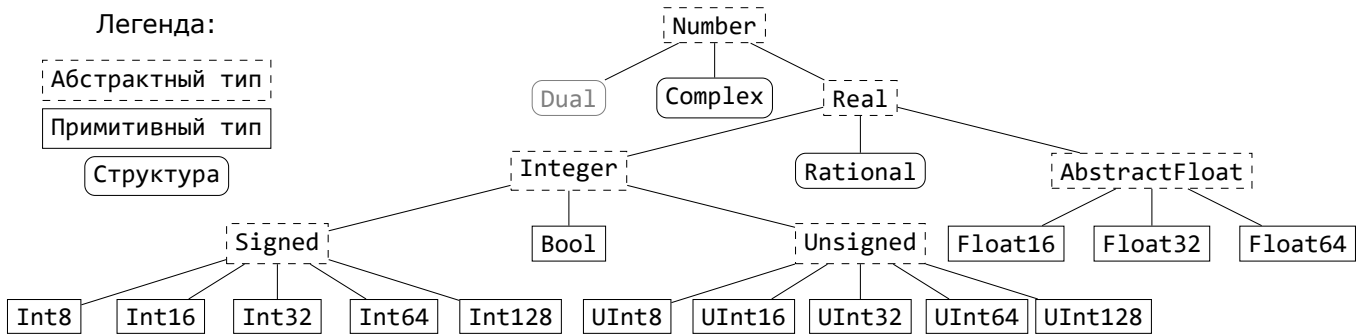}
  \caption{Our Dual type in Julia built-in types hierarchy}
  \label{fig:01}
\end{figure}

Добавление любого пользовательского типа данных в Julia начинается с определения структуры данных. Для дуальных чисел определим структуру
\verb|Dual|:
\begin{minted}{julia}
  struct Dual{T<:Real} <: Number
    "real part"
    x::T
    "imaginary part"
    y::T
  end
\end{minted}

Структура содержит всего два поля: действительную \verb|x| и мнимую \verb|y| части дуального числа. Оба поля структуры должны иметь одинаковый параметрический тип \verb|T|, который должен быть подтипом абстрактного типа \mintinline{julia}{Real}, о чём говорит оператор \mintinline{julia}{<:}. Сам же тип \verb|Dual| является подтипом абстрактного типа \mintinline{julia}{Number}. Таким образом дуальные числа встраиваются в уже существующую иерархию типов (см. рис.~\ref{fig:01}).

Сразу после объявления новой структуры мы можем создать переменные типа Dual, используя конструкторы по умолчанию. Мы обязательно должны задать оба поля структуры в виде аргументов функции Dual:
\begin{minted}{julia}
  z = Dual(1, 2) # одинаковые тип T у аргументов
  Dual{Float64}(1, 2.2) # приведение аргументов к одному типу
\end{minted}

Если тип аргументов одинаков, то явное указание параметра \verb|T| не требуется. Если же  аргументы имеют разный тип, то указание параметрического типа обязательно. В нашем примере оба аргумента приводятся к типу \mintinline{julia}{Float64}.

  \subsection{Перегрузка функций на примере show}

Для возможности распечатывать значения переменных типа Dual, следует перегрузить функцию \verb|show|. Эта функция предназначена для форматированной распечатки данных в стандартный вывод и в REPL. Функции \verb|write|, \verb|print| и \verb|println| используются для вывода минимального текстового представления данных. Из них последние две распечатывают данные только в стандартный поток вывода. Для простых данных можно не делать разницы между форматированным и компактным представлением и перегрузить только \verb|show|. Тогда остальные функции будут использовать \verb|show| в своей работе.

В терминах множественной диспетчеризации перегрузка функции означает добавление нового \emph{метода} к функции. При добавлении метода для функции \verb|show| следует определить её следующим образом:
\begin{minted}{julia}
  Base.show(io::IO, z::Dual) = show(io, z.x, " + ", z.y, "ε")
\end{minted}
Здесь использовался сокращённый синтаксис задания функций. Префикс \verb|Base.| нужен, так как стандартные методы функции \verb|show| находятся в модуле \verb|Base|. Функции из этого модуля доступны для вызова непосредственно без префикса, но при необходимости перегрузки применение \verb|Base.| обязательно.

Стоит отметить, что наш вариант метода \verb|show| упрощён, так как не учитывает частные случаи, например, отрицательную мнимую часть.

Далее мы добавим методы для многих функций из \verb|Base|, для того чтобы наш тип \verb|Dual| стал в достаточной мере функциональным.

  \subsection{Дополнительные конструкторы}

Создадим дополнительные конструкторы. Для этого следует перегрузить конструктор по умолчанию \verb|Dual|.

Всего мы зададим 5 дополнительных конструкторов:

\begin{minted}{julia}
Dual(x::Real, y::Real)= Dual(promote(x, y)...) # 1
Dual{T}(x::Real) where {T<:Real} = Dual{T}(x, 0) # 2
Dual(x::Real) = Dual(promote(x, 0)...) # 3
Dual{T}(z::Dual) where {T<:Real} = Dual{T}(z.x, z.y) # 4
Dual(z::Dual) = Dual(promote(z.x, z.y)...) # 5
\end{minted}

  Первый конструктор позволит нам не указывать всякий раз параметр \verb|T|, если аргументы имеют разный тип. Функция \mintinline{julia}{promote} осуществляет приведение типов переданных ей аргументов к общему типу и возвращает результат в виде кортежа. Постфиксный оператор \verb|...| распаковывает картеж и передаёт его элементы в виде аргументов в функцию-конструктор. В ядре языка определены правила приведения для всех подтипов абстрактного типа \mintinline{julia}|Real|, поэтому теперь конструктор будет корректно работать для любой комбинации аргументов, главное чтобы выполнялось правило \mintinline{julia}|T<:Real|. Например, следующий код сработает корректно:
\begin{minted}{julia}
  Dual(1//2, π) # 0.5 + π*ε
\end{minted}

Мы передали в конструктор рациональное число (тип \mintinline{julia}|Rational|) и встроенную глобальную константу (число $\pi$) типа \mintinline{julia}|Float64|. После чего сработало правило приведения типов и оба аргумента были приведены к более общему типу \mintinline{julia}|Float64|.
 
Второй и третий дополнительные конструкторы позволят не указывать мнимую часть, в случае если она равна нулю:
\begin{minted}{julia}
  Dual{Float32}(1) # 1.0 + 0ε, 2 конструктор
  Dual(1//2) # 1//2 + 0ε, 3 конструктор
\end{minted}

Конструктор №2 является параметрической функцией, которая объявляется с использованием конструкции \mintinline{julia}|where|. Параметр \verb|T| является подтипом абстрактного типа \mintinline{julia}|Real|.  Конструктор №3 работает аналогично конструктору №1. Четвёртый и пятый конструкторы позволяют передавать в качестве аргумента конструктора другое дуальное число.

Для большего удобства также можно создать отдельную константу для мнимой единицы $\iiz$:
\begin{minted}{julia}
  const ε = Dual(0, 1)
\end{minted}

После перегрузки арифметических операций эта константа позволит создавать новые дуальные числа с помощью выражения, максимально близкого к алгебраической их записи:
\begin{minted}{julia}
  z = 1 + 2ε
\end{minted}

  \subsection{Доступ к полям структуры}

Структуры в Julia по умолчанию неизменяемы (immutable), то есть,  один раз создав переменную \verb|z|, мы можем обращаться к полям самой структуры через оператор точка, однако не можем модифицировать значение этих полей:
\begin{minted}{julia}
  @show z.x z.y # можно считывать значение поля
  z.x = 1 # Ошибка! Нельзя изменять.
\end{minted}

Для создания изменяемых (mutable) типов данных предусмотрена структура \mintinline{julia}{mutable struct}, однако для нашего примера более подходящим будет именно неизменяемая структура, так как требование изменяемости накладывает ограничения в производительности, что для числового типа нежелательно.

В Julia нет модификаторов доступа к полям типа \verb|public| или \verb|private|, поэтому поля структуры всегда доступны. Однако хорошим стилем программирования считается инкапсуляция полей структуры и предоставление интерфейса для доступа к ним. В таком стиле выполнен, например, встроенный тип комплексных чисел.

Существуют два способа инкапсуляции полей.  Первый способ заключается в создании специальных интерфейсных функций для доступа к полям. В нашем случае достаточно определить следующие функции:

\begin{minted}{julia}
  Base.real(z::Dual) = z.x
  Base.imag(z::Dual) = z.y
  Base.reim(z::Dual) = (z.x, z.y)
\end{minted}

Такие же функции определены во встроенном модуле Complex.  Далее везде для получения полей структуры следует вызывать только эти функции вместо обращения к полям по именам напрямую. Это позволит разработчикам в дальнейшем производить рефакторинг структуры, например, переименовать или добавлять новые поля. Обратную совместимость при этом легко сохранить, переписав лишь интерфейсные функции. Производительность при этом страдать не будет, так как оптимизирующий JIT-компилятор заменит вызовы этих функций непосредственно на их код, так как они крайне просты (inline functions).

Второй подход заключается в использовании функции \mintinline{julia}{getproperty}, которая появилась начиная с версии 0.7 языка Julia. Перегрузка этой функции позволяет задавать дополнительные имена для обращения к полям структуры. Так, если для нашего примера написать следующий метод:
\begin{minted}{julia}
function Base.getproperty(z::Dual, p::Symbol)
  if p in (:real, :a, :r) # действительная часть
    return getfield(z, :x)
  elseif p in (:imag, :b, :i)  # мнимая часть
    return getfield(z, :y)
  else
    return getfield(z, p)
  end
end
\end{minted}
то мы получим возможность обращаться к действительной и мнимой частям числа $z$ четырьмя разными способами:
\begin{minted}{julia}
  z.x == z.a == z.r == z.real # вернет true
  z.y == z.b == z.i == z.imag # вернет true
\end{minted}

Такой подход более гибок, так как при изменении структуры для обратной совместимости достаточно будет модифицировать лишь один метод \mintinline{julia}{getproperty}, а не все интерфейсные функции, которых может быть много, если структура сложная. Кроме того, пользователю будут доступны все варианты обращения к полям структуры.

Для типа Dual мы применили оба подхода, так как математически обосновано наличие функций \verb|real| --- эквивалента $\Re(z)$, и \verb|imag| --- эквивалента $\Im(z)$.

  \subsection{Унарные функции}

В модуле \verb|Base| определены функции \verb|one| и \verb|zero|, которые возвращают нейтральный элемент по умножению и по сложению соответственно. В случае дуальных чисел нейтралом по умножению является вещественная единица, а нейтралом по сложению --- вещественный ноль.

Для типа Dual  определим следующие однострочные методы:

\begin{minted}{julia}
  Base.one(::Type{Dual{T}}) where T <: Real = Dual(one(T))
  Base.one(z::Dual) = Dual(one(z.x))
  
  Base.zero(::Type{Dual{T}}) where T <: Real = Dual(zero(T))
  Base.zero(z::Dual) = Dual(zero(z.x))
\end{minted}

Первый вариант функций \verb|one| и \verb|zero| принимает в качестве аргумента тип данных и возвращает единицу или ноль этого типа соответственно. Так как для дуальных чисел это $1$ и $0$ из $\mathbb{R}$, то достаточно вернуть дуальное число с нулевой мнимой частью. Вещественная же часть будет $1$ и $0$ параметрического типа \verb|T|. Для всех стандартных типов \verb|T <: Real| методы \verb|one| и \verb|zero| определены в модуле \verb|Base|, чем мы и воспользовались.

Второй вариант функций в качестве аргумента принимает уже конкретный объект типа \verb|Dual| и также возвращает нейтрал с помощью методов из модуля \verb|Base|.

Для комплексных чисел в \verb|Base| определены функции сопряжения \verb|conj|, абсолютного значения \verb|abs| и аргумента \verb|arg|:

\begin{minted}{julia}
  Base.conj(z::Dual) = Dual(z.x, -z.y) 
  Base.abs(z::Dual)  = abs(z.x)
  Base.abs2(z::Dual) = z.x^2
  function Base.arg(z::Dual)
    @assert !iszero(z.x)
    return z.y / z.x
  end
\end{minted}

Для метода \verb|arg| мы предусмотрели проверку на неравенство нулю действительной части дуального числа, так как иначе мы получим деление на ноль. Использование стандартной функции \verb|iszero| позволяет не беспокоиться об учёте погрешностей представления вещественных чисел с помощью чисел с плавающей запятой. Макрос \mintinline{julia}{@assert} выбросит исключение \mintinline{julia}{AssertionError} в случае равенства нулю действительной части.

Функция \verb|inv| определяет обратное число для данного числа
\verb|z|:
\begin{minted}{julia}
  function Base.inv(z::Dual)
    @assert !iszero(z.x)
    return Dual(1/z.x, -z.y/z.x^2)
  end
\end{minted}
Здесь также следует предусмотреть исключение для нулевой действительной части.

  \subsection{Функции сравнения}

В модуле \verb|Base| также определён целый ряд функций, возвращающих
истину, в случае выполнение того или иного условия. Перечислим
некоторые из этих функций:
\begin{itemize}
  \item \verb|isreal| --- число действительное;
  \item \verb|isinteger| --- число целое;
  \item \verb|isfinite| --- число конечно;
  \item \verb|isnan|--- аргумент имеет тип \mintinline{julia}{NaN};
  \item \verb|isinf| --- аргумент имеет тип \mintinline{julia}{Inf};
  \item \verb|iszero| --- число является нейтралом по сложению (ноль);
  \item \verb|isone| --- число является нейтралом по умножению (единица).
\end{itemize}   

Методы для этих функций на случай типа \verb|Dual| один в один
повторяет методы для встроенного типа \verb|Complex|, поэтому здесь мы
не приводим их исходный код.

Кроме унарных функций для дуальных чисел имеет смысл реализовать методы для оператора сравнения \verb|==|. Операторы в Julia ничем не отличаются от функций и добавление методов для них проходит точно также. Единственное отличие заключается в необходимости добавления символа \verb|:| после \verb|Base.|, а так как оператор \verb|==| состоит из двух символов, то необходимо обрамить его  круглыми скобками:
\begin{minted}{julia}
  Base.:(==)(z::Dual, u::Dual) = (real(z) == real(u)) && (imag(z) == imag(u))
\end{minted}

Имеет смысл сравнивать дуальные числа также с действительными, в случае если у дуального числа есть нулевая мнимая часть. Для того чтобы оператор коммутировал для аргументов разного типа, необходимо определить два метода:

\begin{minted}{julia}
Base.:(==)(z::Dual, x::Real) = isreal(z) && real(z) == x
Base.:(==)(x::Real, z::Dual) = isreal(z) && real(z) == x
\end{minted}

  \subsection{Арифметические операции}

Также необходимо перегрузить арифметические операторы, такие как унарные операторы \verb|+| и \verb|-| и бинарные \verb|+|, \verb|-|, \verb|*| и \verb|/|. Реализация операторов \verb|+|, \verb|-| и \verb|*| тривиальна:

\begin{minted}{julia}
  Base.:+(z::Dual) = Dual(+z.x, +z.y)
  Base.:-(z::Dual) = Dual(-z.x, -z.y)
  
  Base.:+(z::Dual, u::Dual) = Dual(z.x + u.x, z.y + u.y)
  Base.:-(z::Dual, u::Dual) = Dual(z.x - u.x, z.y - u.y)
  
  Base.:*(z::Dual, u::Dual) = Dual(z.x * u.x, z.x * u.y + z.y * u.x)
\end{minted}

В случае же оператора \verb|/| следует учесть равенство нулю делителя:
\begin{minted}{julia}
  function Base.:/(z::Dual, u::Dual)
    @assert !iszero(u.x)
    return Dual(z.x/u.x, (z.y*u.x-u.y*z.x)/u.x^2)
  end
\end{minted}

Для бинарных операторов не надо отдельно реализовывать случаи аргументов разного типа, потому что в этом случае должен срабатывать механизм приведения к общему типу с помощью функции \verb|promote|.

  \subsection{Приведение типов}

Для работы механизма приведения типов следует определить правила \emph{приведения типов}. Без этих правил, например, следующая операция будет оканчиваться ошибкой:

\begin{minted}{julia}
  >>Dual(1, 2) + 2
  ERROR: promotion of types Dual{Int64} and Int64 failed to change any arguments
\end{minted}

Эта ошибка возникает из-за того, что нет правила, по которому можно определить общий тип для чисел типа
\mintinline{julia}{Dual{Int64}} и типа \mintinline{julia}{Int64}.

Для определения такого правила нужно добавить метод для функции \verb|promote_rule| из модуля \verb|Base|. В арифметической операции с дуальным числом может участвовать любое число, относящееся к типу \mintinline{julia}{Real}, так как это число можно интерпретировать как дуальное с нулевой мнимой частью:
\begin{minted}{julia}
  Base.promote_rule(::Type{Dual{T}}, ::Type{S}) where {T<:Real, S<:Real} = Dual{promote_type(T, S)}
\end{minted}

Также следует предусмотреть правило, которое будет работать в случае действий с двумя дуальными числами с разными параметрическими типами \verb|T| и \verb|S|. В этом случае нужно найти общий тип для типов \verb|T| и \verb|S|:
\begin{minted}{julia}
  Base.promote_rule(::Type{Dual{T}}, ::Type{Dual{S}}) where {T<:Real, S<:Real} = Dual{promote_type(T, S)}
\end{minted}

  \subsection{Возведение в рациональную степень}

Для возведения числа в степень служит оператор \verb|^|, который также следует перегрузить для целой и рациональной степени. В случае целой степени мы используем формулу
\begin{equation}
  (a + \iiz b)^n = a^n + nba^{n-1}\iiz.
\end{equation}

Следует предусмотреть несколько особых случаев:
\begin{itemize}
  \item если $n=0$, то $z^n = 1$;
  \item если $n=1$, то $z^n = z$;
  \item если $n > 0$, то формула работает для любых $a$ и $b$;
  \item если $n < 0$, то обязательно выполнение условия $a \neq 0$.
\end{itemize}

Если учесть все эти случаи, то получим следующую реализацию:
\begin{minted}{julia}
  function Base.:^(z::Dual, n::Integer)::Dual
    x, y = reim(z)
    if n == 0
      return one(z)
    elseif n == 1
      return z
    elseif n > 1
      return Dual(x^n, n*y*x^(n-1))
    else #n < 0
      if iszero(x)
        throw(DomainError(:x, "negative exponentiation is only defined for real(z)!= 0"))
      else
        return Dual(x^n, n*y*x^(n-1))
      end
    end
  end
\end{minted}

Если действительная часть дуального числа равна нулю (\verb|iszero(x)|) и при этом $n<0$, то функция выбрасывает исключение \mintinline{julia}{DomainError} (выход за область допустимых значений).

Для рациональной степени используется более сложная формула:
\begin{equation}
  (a + \iiz b)^{\frac{n}{m}} = a^{\frac{n}{m}}
  \qty(1 + \iiz\frac{n}{m}\frac{b}{a})
  = a^{\frac{n}{m}} +
  \iiz\frac{n}{m}ba^{\frac{n}{m} - 1},
\end{equation}
для которой следует предусмотреть случаи чётного и нечётного $m$. При нечётном $m$ в область допустимых значений входят $a < 0$, а в случае чётного $m$ следует ограничиться только $a \geqslant 0$:
\begin{minted}{julia}
  function Base.:^(z::Dual, q::Rational)::Dual
    x, y = reim(z)
    n, d = numerator(q), denominator(q)
    if n == 0
      return one(z)
    elseif isodd(d)
      return Dual(x^q, q*y*x^(q-1))
    else
      if x < 0
        throw(DomainError(:x, "even radical for dual number z is only defined for real(z)>=0"))
      else
        return Dual(x^q, q*y*x^(q-1))
      end
    end
  end
\end{minted}

Имеет смысл отдельно перегрузить функции для квадратного и кубических корней \verb|sqrt| и \verb|cbrt|:
\begin{minted}{julia}
  function Base.sqrt(z::Dual)::Dual
    x, y = reim(z)
    if x < 0
      throw(DomainError(:x, "sqrt for dual number z is only defined for real(z)>=0"))
    else
      return Dual(√x, y/(2*√x))
    end
  end
  
  Base.cbrt(z::Dual) = Dual(cbrt(z.x), z.y*cbrt(z.x)/(3z.x))
\end{minted}

  \subsection{Элементарные функции}

Элементарные функции вычисляются по формуле $f(a + \iiz b) = f(a) + f'(a)b\iiz$. Следует учитывать, что многие из них не определены для случая нулевой действительной части числа $z = a + \iiz b$:
\begin{minted}{julia}
  function Base.exp(z::Dual)
    e = exp(z.x)
    return Dual(e, e * z.y)
  end

  Base.log(z::Dual) = Dual(log(z.x), z.y/z.x)
  Base.log(b, z::Dual) = Dual(log(b, z.x), (z.y/z.x) * log(z.x))

  #======== Trigonometric ========#

  Base.sin(z::Dual) = Dual(sin(z.x), +z.y*cos(z.x))
  Base.cos(z::Dual) = Dual(cos(z.x), -z.y*sin(z.x))

  Base.tan(z::Dual) = Dual(tan(z.x),  z.y/cos(z.x)^2)
  Base.cot(z::Dual) = Dual(cot(z.x), -z.y/sin(z.x)^2)

  Base.asin(z::Dual) = Dual(asin(z.x),  z.y/sqrt(1-z.x^2))
  Base.acos(z::Dual) = Dual(acos(z.x), -z.y/sqrt(1-z.x^2))

  Base.atan(z::Dual) = Dual(atan(z.x),  z.y/(1+z.x^2))
  Base.acot(z::Dual) = Dual(acot(z.x), -z.y/(1+z.x^2))

  #======== Hyperbolic ========#

  Base.sinh(z::Dual) = Dual(sinh(z.x), z.y*cosh(z.x))
  Base.cosh(z::Dual) = Dual(cosh(z.x), z.y*sinh(z.x))

  Base.tanh(z::Dual) = Dual(tanh(z.x), z.y/cosh(z.x)^2)
  Base.coth(z::Dual) = Dual(coth(z.x), z.y/sinh(z.x)^2)
\end{minted}
\section{Результаты}
\label{sec:results}

  В работе описана предварительная реализация дуальных комплексных
  чисел и основных операций над ними на языке Julia.

\section{Обсуждение}
\label{sec:discussion}

Реализация дуальных комплексных чисел на Julia полностью опирается на механизм множественной диспетчеризации. Таким образом мы не только реализовали некоторый набор операций над дуальными числами, но и продемонстрировали мощность этого механизма. 
Также следует отметить, что в отличие от реализации типа \verb|Dual| в пакете автоматического дифференцирования \verb|ForwardDiff|~\cite{L_Julia:2020:ForwardDiff}, предложенная нами реализация более чистая. Например, в вышеупомянутом пакете в качестве коэффициента перед дуальной комплексной единицей может выступать обычное комплексное число. Понятно, что это вызвано спецификой применения дуальных чисел в этом пакете для автоматического дифференцирования. Но данного типа числа скорее относятся к кватернионам~\cite{hamilton:1866:elements-quaternions,hamilton:1866:elements-quaternions,yefremov:2005:quaternions-biquaternions}, нежели к собственно комплексным числам.

\section{Заключение}
\label{sec:conclusion}

В работе был сделан прототип реализации обобщённых комплексных чисел на языке Julia, а именно реализация дуальных комплексных чисел. Наличие механизма множественной диспетчеризации в языке Julia очень упрощает реализацию новых числовых типов в рамках существующей инфраструктуры языка программирования. Авторы предполагают в дальнейшем расширить данный прототип до полной реализации обобщённых комплексных чисел и обобщённых кватернионов.

\def\acknowledgmentsname{Благодарности}

\begin{acknowledgments}

  Публикация подготовлена
  при финансовой поддержке РФФИ в рамках научного проекта
\textnumero~19-01-00645.

\end{acknowledgments}

  \ifthenelse{\boolean{xetex} \OR \boolean{luatex}}
  {\bibliographystyle{ugost2008l}}
  {\bibliographystyle{gost2008l}}

\bibliography{bib/dual-numbers/cite}